\documentclass[preprint,12pt]{elsarticle}
\usepackage[colorlinks=true, urlcolor=blue, citecolor=blue]{hyperref}
\usepackage{amssymb}
\usepackage{placeins}
\usepackage{rotating}
\usepackage[table]{xcolor}
\usepackage{lscape}
\usepackage{longtable}
\newcolumntype{L}{>{\centering\arraybackslash}m{6cm}}
\newcolumntype{M}{>{\centering\arraybackslash}m{2.5cm}l}
\newcolumntype{C}[1]{>{\centering\arraybackslash}p{#1}}

\biboptions{sort&compress}

\journal{International Journal of Medical Informatics}

\begin{document}

\begin{frontmatter}

\title{Stress Monitoring Using Low-Cost Electroencephalogram Devices: A Systematic Literature Review}

\author[inst1]{Gideon Vos}

\affiliation[inst1]{organization={College of Science and Engineering, James Cook University},
            addressline={James Cook Dr}, 
            city={Townsville},
            postcode={4811}, 
            state={QLD},
            country={Australia}}

\author[inst1]{Maryam Ebrahimpour}
\author[inst3]{Liza van Eijk}
\author[inst2]{Zoltan Sarnyai}
\author[inst1]{Mostafa Rahimi Azghadi}

\affiliation[inst2]{organization={College of Public Health, Medical, and Vet Sciences, James Cook University},
            addressline={James Cook Dr}, 
            city={Townsville},
            postcode={4811}, 
            state={QLD},
            country={Australia}}

\affiliation[inst3]{organization={College of Health Care Sciences, James Cook University},
            addressline={James Cook Dr}, 
            city={Townsville},
            postcode={4811}, 
            state={QLD},
            country={Australia}}

\begin{abstract}
\paragraph{Introduction}
Low-cost, consumer grade wearable health monitoring devices are increasingly being used for mental health related studies  including stress. While cortisol response magnitude remains the gold standard indicator for stress assessment, a growing number of studies have started to use low-cost EEG devices and wrist-based wearable device monitors as primary recorders of biomarker data. Low-cost EEG devices differ in technical sophistication, and more importantly number of sensors available, and the location of these sensors according to the 10-20 System of Electrode Placement, complicating the reproducibility of reported study results.

\paragraph{Objective}
The aim of this review is to provide an overview of the growing use of low-cost EEG devices and machine learning methods for brain function assessment, specifically stress. We also shed light on the machine learning techniques commonly employed for stress research when using EEG brain function data, with an emphasis on the reproducibility of the results reported.  

\paragraph{Methods}
This study reviewed published works contributing and/or using EEG devices for detecting stress and their associated machine learning methods. The electronic databases of Google Scholar, ScienceDirect, Nature and PubMed were searched for relevant articles and a total of 60 articles were identified and included in the final analysis. The reviewed works were synthesized into three categories of low-cost EEG device usage for stress research, machine learning techniques applied to EEG data for measuring stress, and future research directions. For the machine learning studies reviewed, we provide an analysis of their approach to results validation and reproducibility. The quality assessment of the included studies was conducted in accordance with the IJMEDI checklist.

\paragraph{Results}
A number of studies were identified where low-cost EEG devices were utilized to record brain function during phases of stress and relaxation. These studies generally reported a high predictive accuracy rate, verified using a number of different machine learning validation methods and statistical approaches. Of these studies, only 50\% reported health screening prior to experimentation and a further 60\% can be considered low-powered studies based on the small number of test subjects used during experimentation. In addition, we discuss that previous works focused on a single health condition (stress) with little consideration for other potential underlying mental health conditions and how it may affect the study results reported.

\paragraph{Conclusion}
Low-cost consumer grade wearable devices including EEG and wrist-based monitors are increasingly being used in stress-related studies. Standardization of EEG signal processing and importance of sensor location still requires further study, and research in this area will continue to provide improvements as more studies become available. 

\end{abstract}

\begin{graphicalabstract}
\begin{center}
  \makebox[\textwidth]{\includegraphics[width=\paperwidth]{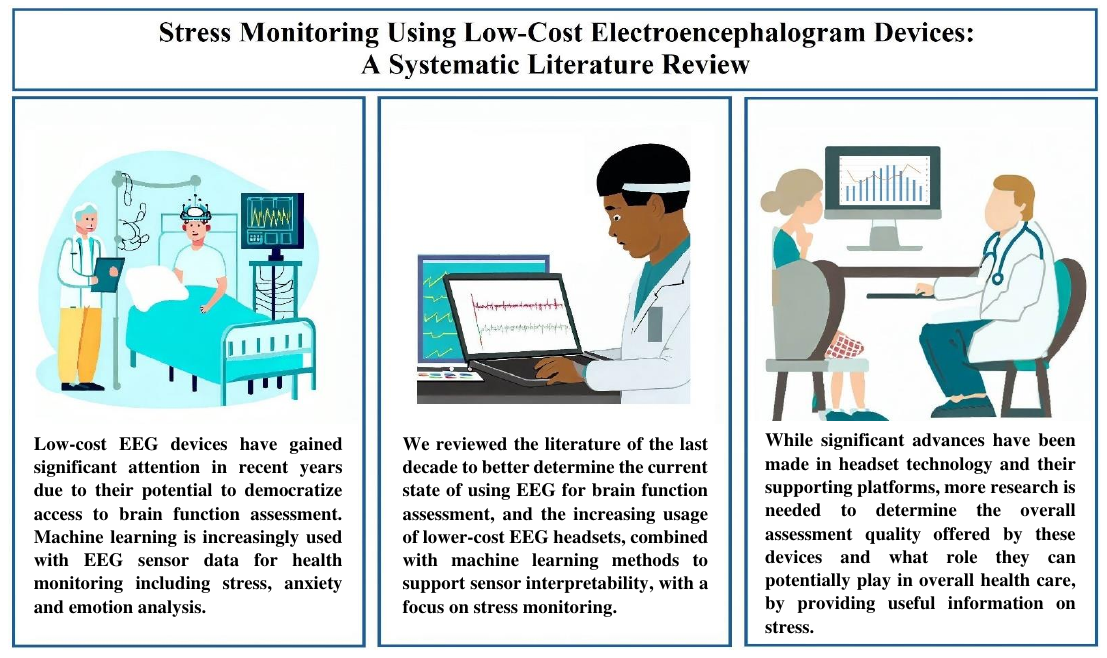}}
\end{center}
\end{graphicalabstract}

\begin{highlights}
\item This paper provides a review of the current state of utilizing ready to use, wearable EEG devices and machine learning methods for stress measurement, performed in accordance with the IJMEDI checklist.
\item Our review of EEG devices is provided by analyzing and synthesizing the literature based on machine learning methods used, study quality and reproducibility of results, with a further emphasis on the growing use of low-cost EEG devices.
\item We show that the combination of low-cost EEG devices with machine learning methods provide a viable alternative to the use of more intrusive medical grade devices, with a growing number of studies incorporating the use of wrist-based wearable health monitors for recording additional biomarkers.

\end{highlights}


\end{frontmatter}


\section{Introduction}

\noindent Stress can be defined as any type of change that causes physical, emotional, or psychological strain. These changes in the environment can trigger a cascade of biological responses in the brain and body \cite{McEwen1998}, known as the stress response, which helps the organism adapt to the dynamically changing external and internal environment. This adaptation is achieved through the mobilization of energy and its appropriate redistribution to organs that serve the adaptational response. While the stress response is adaptive and beneficial in the short term, long-term exposure to stress can have detrimental effects on the body, including an increased risk of developing metabolic, cardiovascular, and mental disorders, which can significantly compromise quality of life and shorten life expectancy \cite{McEwen2007}.\\

\noindent The rise in mental health issues \cite{McGrath2023} and the exacerbating effects of the COVID-19 pandemic further highlights the importance of addressing stress and its impact on mental health. Understanding the physiological and psychological changes that occur in the body during stress response can inform the development of effective interventions to mitigate the negative effects of stress on mental health.\\

\noindent The stress response is initiated by the brain as the different sensory cortical areas receive and process information from the changing environment and distribute them to a variety of cortical areas to initiate the appropriate response. Therefore, stress changes neuronal activity in the whole brain, which can be detected by measuring the electrical activation of the cerebral cortex by EEG \cite{Hendry2024}. 

\begin{figure}[h!]
\centering
\includegraphics[width=\textwidth]{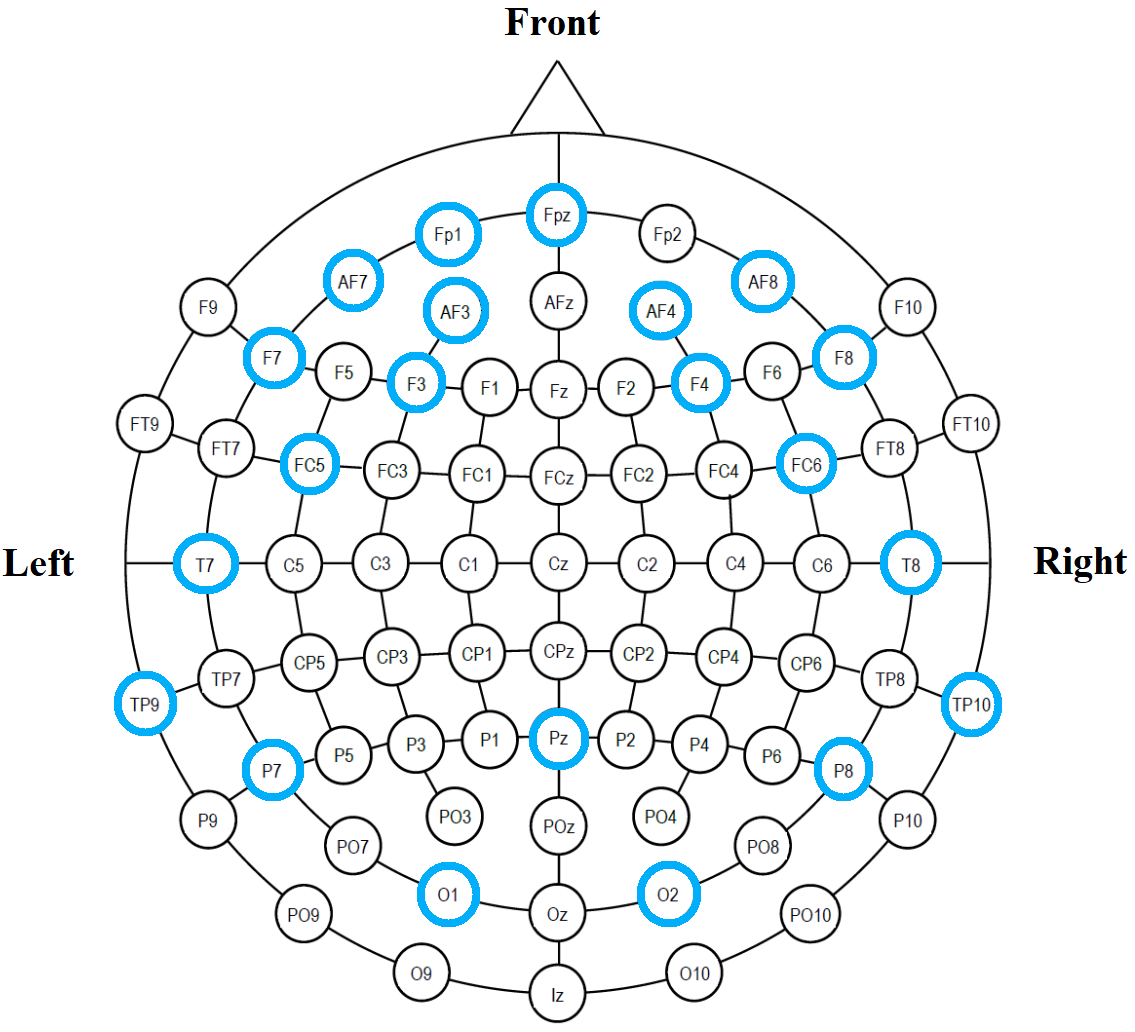}
\caption{\label{fig:figure1}The 10-20 international standard of EEG sensor placement, with sensor placements supported by devices included in this review indicated in blue..}%
\end{figure}

\noindent EEG is a widely used technique to estimate changes in neurophysiological activity associated with external stimuli and/or with the performance of specific tasks \cite{Giannakakis2019}. This activity is measured via the electrical potential generated by the asynchronous ﬁring of neurons \cite{Zhang_2023} in the nervous system \cite{Halim2020} using electrodes placed on the scalp according to the 10-20 international standard \cite{Jasper1958} (Figure \ref{fig:figure1}). The 10-20 system was first presented at the 1957 Brussels IV International EEG Congress by Herbert Jasper, to standardize the method of EEG placement. The numbers `10' and `20' refer to the distances between adjacent electrodes, which are either 10\% or 20\% of the total distance (front-back or right-left) of the skull.\\

\FloatBarrier

\begin{figure}[h!]
\centering
\includegraphics[width=\textwidth]{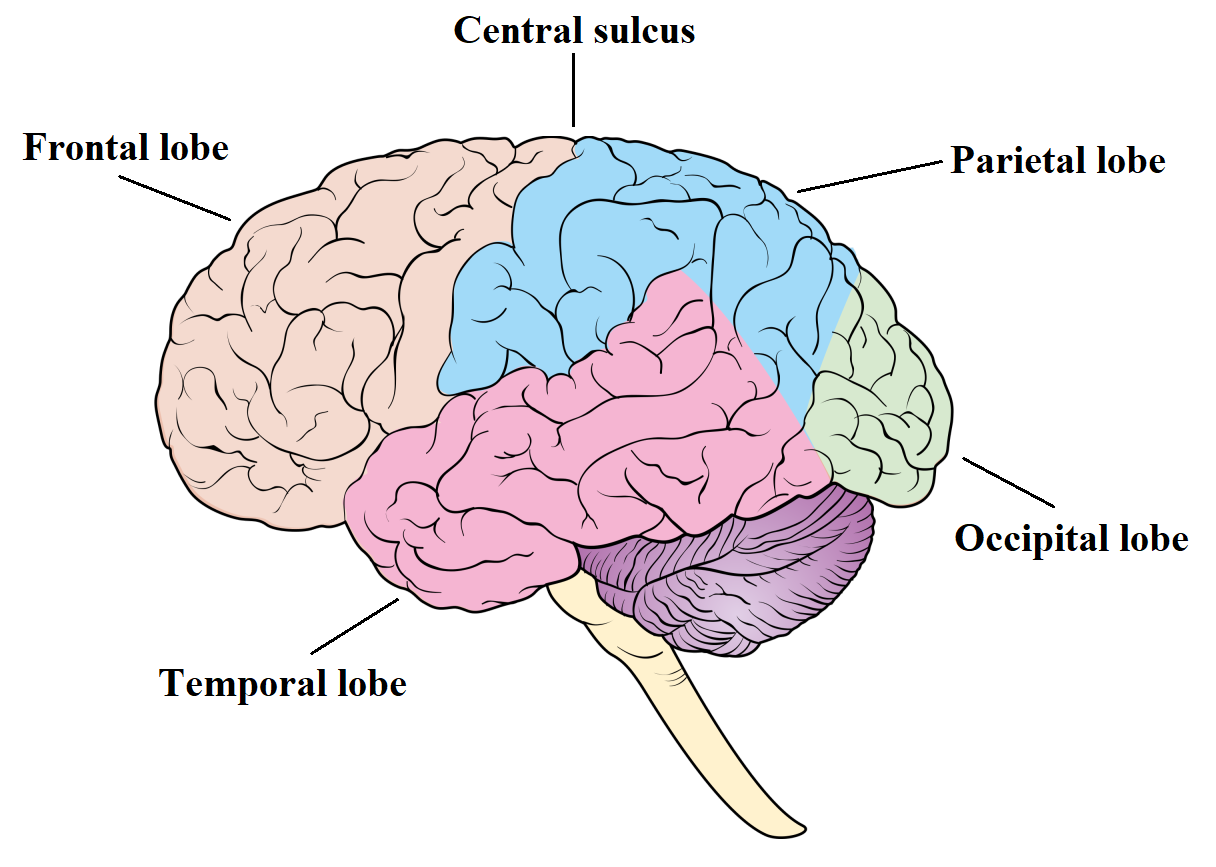}
\caption{\label{fig:figure2}Anatomy of the human brain \cite{physiopedia2024}.}%
\end{figure}

\noindent The 10-20 system is based on the relationship between the location of an electrode and the underlying area of the cerebral cortex (Figure \ref{fig:figure2}). Each site has a letter (to identify the lobe) and a number or another letter to identify the hemisphere location. In 1985, an extension to the original 10-20 system was proposed involving an increase in the number of electrodes from 21 to 74 \cite{Chatrian1988}. \\

\noindent EEG waveforms may be characterized based on their location, amplitude, frequency, morphology, continuity (rhythmic, intermittent or continuous), synchrony, symmetry, and reactivity. However, the most frequently used method to classify EEG waveforms is by the frequency band, with the most commonly studied waveforms being delta (0.5 to 4Hz), theta (4 to 7Hz), alpha (8 to 12Hz), sigma (12 to 16Hz), beta (13 to 30Hz) and gamma ($>$ 30Hz) \cite{Nayak2023}. Common brain states typically associated with different EEG waveform frequencies include concentration (gamma), anxiety dominant, active, external attention, relaxed (beta), very relaxed, passive attention (alpha), deeply relaxed, inward focused (theta) and sleep (delta) \cite{Abhang2016}. \\

\FloatBarrier

\noindent Prior studies \cite{Siirtola2020, Gjoreski2017, Samson2020} have utilized physiological measures including heart rate (HR), electrodermal activity (EDA) and salivary cortisol to assess the sympathetic nervous system and Hypothalamic-Pituitary-Adrenal (HPA) axis response to stress. These measures can be influenced by multiple factors including mental stress, and can further be affected by circadian rhythm and physical activity. Functional neuroimaging techniques including EEG have also previously been used to assess the brain's response to stress \cite{Giannakakis2019, Katmah2021, Vanhollebeke2022} by directly or indirectly measuring brain activity. EEG offers high temporal resolution, at the cost of requiring trained neurophysiologists to aid in the interpretation of results.\\

\begin{figure}[h!]
\centering
\includegraphics[width=\textwidth]{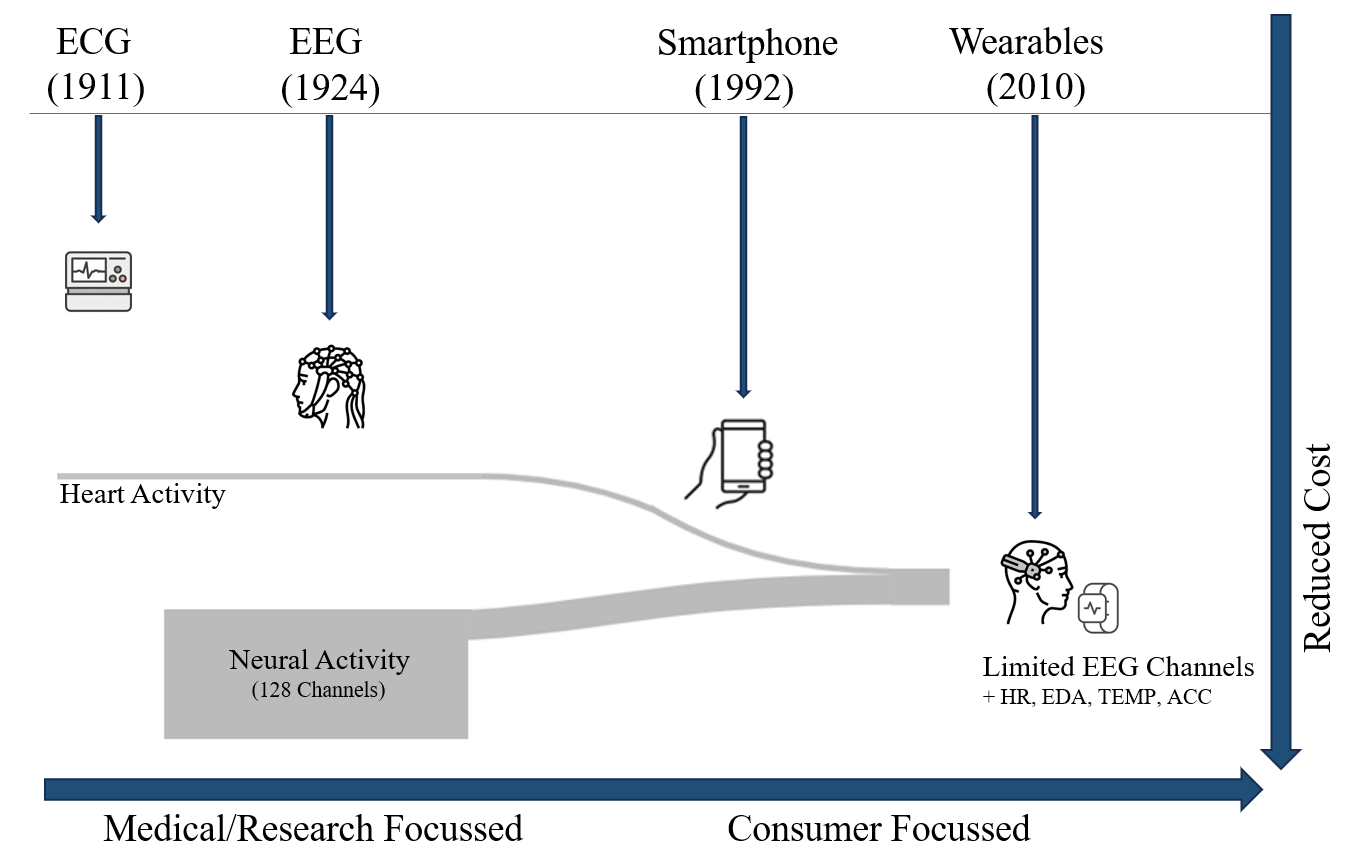}
\caption{\label{fig:figure3}Technological timeline of electronic healthcare monitoring.}%
\end{figure}

\noindent Although the assessment of cortisol response continues to be the primary indicator for evaluating stress \cite{Boucher2019}, an increasing number of studies are now employing affordable, non-intrusive wearable health monitors and wireless EEG devices as the main tools for recording biomarker data \cite{Lee2020, Park2020, Aspiotis2022, Affanni2022} that may correlate with stress. This is likely driven by the increasing sophistication and miniaturization of device components at an ever reducing cost. Figure \ref{fig:figure3} provides a visual timeline of the evolution of electronic healthcare monitoring, illustrating the merging of traditional ECG with EEG, resulting in more modular and mobile devices with a wide range of sensors, albeit with fewer EEG channels.\\

\noindent Most wrist-based wearable health monitors previously utilized in stress research record at least heart rate and electrodermal activity \cite{Vos2023}. In contrast, low-cost EEG devices differ in the number of available sensors and sensor locations according to the 10-20 System of Electrode Placement \cite{Jasper1958}, complicating the reproducibility of reported study results.\\

\noindent A number of previous survey articles have studied the topics of stress detection using EEG \cite{Newson2019, Katmah2021, Vanhollebeke2022, Giannakakis2019}. Katmah \emph{et al.} \cite{Katmah2021} reviewed existing EEG signal analysis methods for assessing mental stress, and concluded that variations in data analysis methods resulted in several contradictory results with regards to frequency band importance. These variations could be due to the study differences in experimental protocols, stressors employed, preprocessing of signals and choice of machine learning algorithms utilized. Newson \emph{et al.} \cite{Newson2019} reviewed 184 studies where EEG was used to study a varying number of mental disorders, and found differences in frequency bands in the resting state condition (eyes open and closed) across a spectrum of psychiatric disorders. They further cautioned against the interpretation of results that considered a single disorder in isolation when analyzing results. While their study focussed on mental disorders, the question arises whether similar differences may be relevant to stress-related EEG studies.\\

\noindent None of the prior reviews specifically addressed EEG hardware and more specifically, the use of low-cost EEG devices for stress research. While Giannakakis \emph{et al.} \cite{Giannakakis2019} reviewed sensor placement and relative importance, and Katmah \emph{et al.} \cite{Katmah2021} addressed the use of machine learning algorithms and signal processing techniques, neither highlighted sensor placement or sensor limitations of low-cost EEG devices when used for stress research, and how this could potentially affect machine learning model performance and the reproducibility of the results reported in stress-related studies.\\

\noindent Towards addressing these questions, we first explore the current state of stress detection and measurement using EEG, with a focus on low-cost EEG devices. We further explore the available public datasets built using sensor data recorded from low-cost EEG devices that could enable future stress research. We review the machine learning approaches employed, and detection accuracy scores attained using low-cost EEG devices. Finally, we discuss the limitations of using such devices combined with machine learning techniques for stress-related studies and suggest several future research directions. \\

\section{Methods}
\subsection{Research questions}
\noindent The main aim of this work is to provide an overview of the use of EEG devices for stress detection and measurement when combined with machine learning techniques, with an emphasis on low-cost wearable EEG devices retailing for below USD1000.00. To assist in the assessment of the quality of the included literature, the IJMEDI checklist was utilized. Thus, our research questions can be formulated as follows:

\begin{itemize}
\item {RQ1: Which low-cost EEG devices are predominantly being used for stress research, and what results are being reported?} 
\item {RQ2: Which scalp sensor locations and frequency bands are predominantly used for stress research, and are these physically supported by current low-cost EEG devices?}
\item {RQ3: Which machine learning methods are being utilized in low-cost EEG studies for stress research, and how are the results from these studies validated?} 
\end{itemize}

\noindent Answering these research questions will aid in gaining a better understanding of the feasibility and potential of using lower-cost EEG devices for stress monitoring, while providing guidance for the most appropriate machine learning algorithms to employ for stress prediction.\\

\subsection{Search strategy}
\begin{figure}[h!]
\centering
\includegraphics[width=\textwidth]{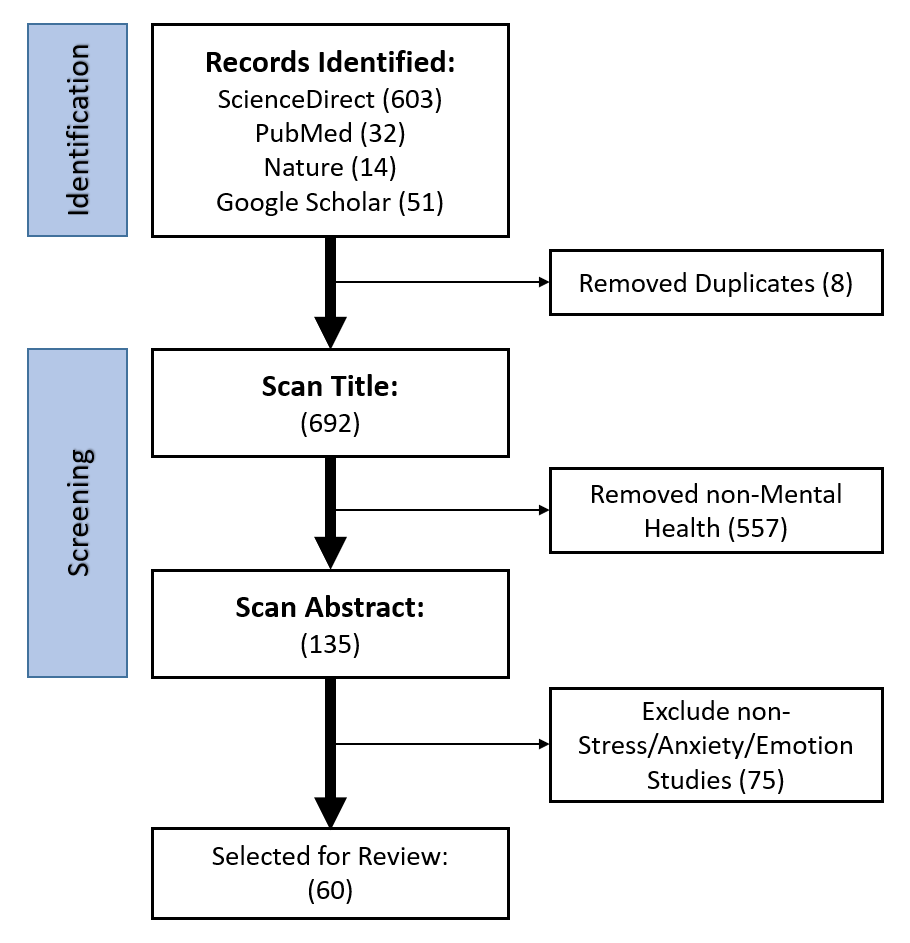}
\caption{\label{fig:figure4}Article screening process and the intermediate counts.}%
\end{figure}

\noindent We reviewed key published works (Figure \ref{fig:figure4}) between 2013 and 2023 on acute stress assessment using EEG signals, and additionally, recorded using low-cost EEG devices. The electronic databases of Google Scholar, ScienceDirect, Nature and PubMed were searched for relevant articles using the keywords \emph{EEG AND stress AND (wearable OR machine learning)}, in title or abstract, and a total of 700 papers were found. Duplicates were identified, and 8 were found and removed, leaving the number of considered papers for the subsequent phases at 692. Titles and abstracts were scanned and irrelevant papers were excluded.\\

\begin{landscape}
\begin{figure}[h!]
\centering
\includegraphics[width=22cm]
{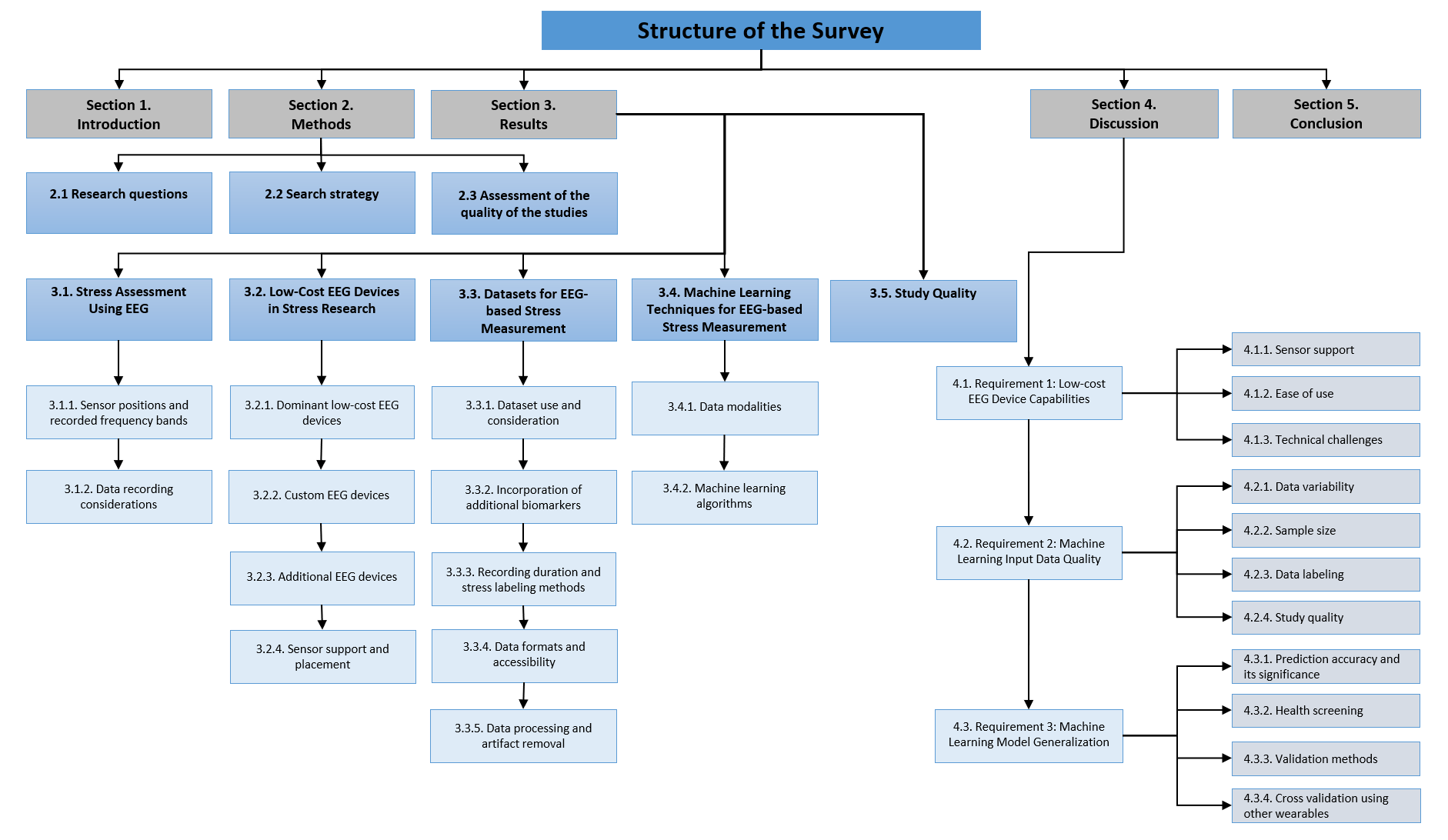}
\caption{\label{fig:figure5}Survey structure.}%
\end{figure}
\end{landscape}

\noindent A small number of papers, in which the focus was stress in animals were excluded. Studies where the main focus was mental health conditions other than stress, including schizophrenia, bipolar disorder and major depression disorder were also excluded. Studies utilizing medical-grade EEG devices were not excluded at this stage. As a result, a total of 60 papers were chosen for the systematic review process, to assist in answering the aforementioned research questions. The chosen papers were then 
grouped by the high-level topics of: \emph{RQ1: Stress Assessment Using EEG}, \emph{RQ2: Low-Cost EEG Devices}, \emph{RQ3: Available Datasets for EEG-based Stress Measurement} and \emph{RQ3: Machine Learning Techniques for EEG-based Stress Measurement}. Table \ref{tab:papersreviewed} lists, in chronological order, the papers included in this review. We further categorize these papers by study type, i.e. Original Research or Review, and domain, i.e. stress or other mental states. A single study \cite{Koelstra2012} that pre-dates the search criteria (2012) was included in this review due to its popular use as an open EEG dataset and high citation count. To answer the research questions systematically, we have structured our paper as shown in Fig. \ref{fig:figure5} to cover all the essential topics around utilizing low-cost EEG devices for stress detection using machine learning.\\

\begin{landscape}
\fontsize{10}{11}\selectfont
\begin{longtable}{ p{1cm} p{1cm} p{10cm} p{2.2cm} p{3cm} }
\caption{Studies included in this review.} \\
\hline\hline
\textbf{Paper} & \textbf{Year} & \multicolumn{1}{l}{\textbf{Title}} & \textbf{Study Type}  & \textbf{Domain}  \\
\hline
\hline\endhead  
\hline\endfoot  

\rowcolor[rgb]{0.753,0.753,0.753} \cite{Koelstra2012} & 2012 & DEAP: A database for emotion analysis using physiological signals                                                                                                                                                & Research            & Emotion              \\
\cite{Borghini2014}       & 2014                              & Measuring neurophysiological signals in aircraft pilots and car drivers for the assessment of mental workload, fatigue and drowsiness                                                                            & Research            & Stress               \\
\rowcolor[rgb]{0.753,0.753,0.753} \cite{Hou2015}            & 2015                              & EEG based stress monitoring                                                                                                                                                                                      & Research            & Stress               \\
\cite{Lopez2015}          & 2015                              & Automated identification of abnormal adult EEGs                                                                                                                                                                  & Research            & Normal/Abnormal      \\
\rowcolor[rgb]{0.753,0.753,0.753} \cite{Zheng2015}          & 2015                              & Investigating critical frequency bands and channels for EEG-based emotion recognition with deep neural networks                                                                                                  & Research            & Emotion              \\
\cite{Giannakakis2015}    & 2015                              & Detection of stress/anxiety state from EEG features during video watching                                                                                                                                        & Research            & Stress               \\
\rowcolor[rgb]{0.753,0.753,0.753} \cite{Minguillon2016}     & 2016                              & Stress assessment by prefrontal relative gamma                                                                                                                                                                   & Research            & Stress               \\
\cite{UmarSaeed2018}      & 2018                              & Selection of neural oscillatory features for human stress classification with single channel EEG headset                                                                                                         & Research            & Stress               \\
\rowcolor[rgb]{0.753,0.753,0.753} \cite{Babayan2019}        & 2019                              & A mind-brain-body dataset of MRI, EEG, cognition, emotion, and peripheral physiology in young and old adults                                                                                                     & Research            & Emotion              \\
\cite{Giannakakis2019}       & 2019                              & Review on psychological stress detection using biosignals & Review            & Stress              \\
\rowcolor[rgb]{0.753,0.753,0.753} \cite{Baghdadi2019}       & 2019                              & DASPS: A database for anxious states based on a psychological stimulation                                                                                                                                        & Research            & Anxiety              \\
\cite{Newson2019}         & 2019                              & EEG frequency bands in psychiatric disorders: A review of resting state studies                                                                                                                                  & Review              & Various              \\
\rowcolor[rgb]{0.753,0.753,0.753} \cite{Ahn2019}            & 2019                              & A Novel wearable EEG and ECG recording system for stress assessment                                                                                                                                              & Research            & Stress               \\

\cite{Arsalan2019}            & 2019                              & Classification of perceived human stress using physiological signals                                                                                                                                              & Research            & Stress               \\

\rowcolor[rgb]{0.753,0.753,0.753} \cite{Lee2020}            & 2020                              & Stress monitoring using multimodal bio-sensing headset                                                                                                                                                           & Research            & Stress               \\
\cite{Park2020}           & 2020                              & K-EmoCon, a multimodal sensor dataset for continuous emotion recognition in naturalistic conversations                                                                                                           & Research            & Emotion              \\
\rowcolor[rgb]{0.753,0.753,0.753} \cite{Saeed2020}          & 2020                              & EEG based classification of long-term stress using psychological labeling                                                                                                                                        & Research            & Stress               \\
\cite{TuerxunWaili2020}   & 2020                              & Stress recognition using Electroencephalogram (EEG) signal                                                                                                                                                       & Research            & Stress               \\
\rowcolor[rgb]{0.753,0.753,0.753} \cite{Parent2020}         & 2020                              & PASS: A multimodal database of physical activity and stress for mobile passive body/brain-computer interface research                                                                                           & Research            & Stress               \\
\cite{Pernice2020}        & 2020                              & Multivariate correlation measures reveal structure and strength of brain-body physiological networks at rest and during mental stress                                                                            & Research            & Stress               \\
\rowcolor[rgb]{0.753,0.753,0.753} \cite{Alakus2020}         & 2020                              & Database for an emotion recognition system based on EEG signals and various computer games – GAMEEMO                                                                                                             & Research            & Emotion              \\
\cite{Halim2020}          & 2020                              & On identification of driving-induced stress using electroencephalogram signals: A framework based on wearable safety-critical scheme and machine learning                                                        & Research            & Stress               \\
\rowcolor[rgb]{0.753,0.753,0.753} \cite{Hu2020}             & 2020                              & Detecting fatigue in car drivers and aircraft pilots by using non-invasive measures: The value of differentiation of sleepiness and mental fatigue                                                               & Review              & Stress               \\
\cite{Mohanavelu2020}     & 2020                              & Dynamic cognitive workload assessment for fighter pilots in simulated fighter aircraft environment using EEG                                                                                                     & Research            & Stress               \\
\rowcolor[rgb]{0.753,0.753,0.753} \cite{Peterson2020}       & 2020                              & A feasibility study of a complete low-cost consumer-grade brain-computer interface system                                                                                                                        & Review              & Various              \\
\cite{Zhang2020}          & 2020                              & Emotion recognition using multi-modal data and machine learning techniques: A tutorial and review                                                                                                                & Review              & Emotion              \\
\rowcolor[rgb]{0.753,0.753,0.753} \cite{Saeed2020}          & 2020                              & EEG based classification of long-term stress using psychological labeling                                                                                                                                        & Research            & Stress               \\
\cite{TuerxunWaili2020}   & 2020                              & Stress recognition using electroencephalogram (EEG) signal                                                                                                                                                       & Research            & Stress               \\
\rowcolor[rgb]{0.753,0.753,0.753} \cite{Parent2020}         & 2020                              & PASS: A multimodal database of physical activity and stress for mobile passive body/ brain-computer interface research                                                                                           & Research            & Stress               \\
\cite{Pernice2020}        & 2020                              & Multivariate correlation measures reveal structure and strength of brain-body physiological networks at rest and during mental stress                                                                            & Research            & Stress               \\
\rowcolor[rgb]{0.753,0.753,0.753} \cite{Alakus2020}         & 2020                              & Database for an emotion recognition system based on EEG signals and various computer games – GAMEEMO                                                                                                             & Research            & Emotion              \\
\cite{Halim2020}          & 2020                              & On identification of driving-induced stress using electroencephalogram signals: A framework based on wearable safety-critical scheme and machine learning                                                        & Research            & Stress               \\
\rowcolor[rgb]{0.753,0.753,0.753} \cite{Hu2020}             & 2020                              & Detecting fatigue in car drivers and aircraft pilots by using non-invasive measures: The value of differentiation of sleepiness and mental fatigue                                                               & Review              & Stress               \\
\cite{Mohanavelu2020}     & 2020                              & Dynamic cognitive workload assessment for fighter pilots in simulated fighter aircraft environment using EEG                                                                                                     & Research            & Stress               \\
\rowcolor[rgb]{0.753,0.753,0.753} \cite{Peterson2020}       & 2020                              & A feasibility study of a complete low-cost consumer-grade brain-computer interface system                                                                                                                        & Review              & Various              \\
\cite{Zhang2020}          & 2020                              & Emotion recognition using multi-modal data and machine learning techniques: A tutorial and review                                                                                                                & Review              & Emotion              \\
\rowcolor[rgb]{0.753,0.753,0.753} \cite{Katmah2021}         & 2021                              & A review on mental stress assessment methods using EEG signals                                                                                                                                                   & Review              & Stress               \\
\cite{Sundaresan2021}     & 2021                              & Evaluating deep learning EEG-based mental stress classification in adolescents with autism for breathing entrainment BCI                                                                                         & Research            & Stress            \\   
\rowcolor[rgb]{0.753,0.753,0.753} \cite{AlSaggaf2021}       & 2021                              & Performance evaluation of EEG based mental stress assessment approaches for wearable devices                                                                                                                     & Research            & Stress               \\
\cite{Hag2021}            & 2021                              & Enhancing EEG-based mental stress state recognition using an improved hybrid feature selection algorithm                                                                                                         & Research            & Stress               \\
\rowcolor[rgb]{0.753,0.753,0.753} \cite{Arsalan2021}        & 2021                              & Human stress classification during public speaking using physiological signals                                                                                                                                   & Research            & Stress               \\
\cite{Chae2021}           & 2021                              & Relationship between rework of engineering drawing tasks and stress level measured from physiological signals                                                                                                    & Research            & Stress               \\
\rowcolor[rgb]{0.753,0.753,0.753} \cite{DiazPiedra2021}     & 2021                              & Monitoring army drivers’ workload during off-road missions: An experimental controlled field study                                                                                                               & Research            & Stress               \\
\cite{Sakalle2021}        & 2021                              & A LSTM based deep learning network for recognizing emotions using wireless brainwave driven system                                                                                                               & Research            & Emotion              \\
\rowcolor[rgb]{0.753,0.753,0.753} \cite{Tan2021}            & 2021                              & NeuroSense: Short-term emotion recognition and understanding based on spiking neural network modelling of spatio-temporal EEG patterns                                                                           & Research            & Emotion              \\
\cite{Wan2021}            & 2021                              & A review on transfer learning in EEG signal analysis                                                                                                                                                             & Review              & Various              \\
\rowcolor[rgb]{0.753,0.753,0.753} \cite{Yeom2021}           & 2021                              & Psychological and physiological effects of a green wall on occupants: A cross-over study in virtual reality                                                                                                      & Research            & Stress               \\
\cite{Zhang2021}          & 2021                              & Exploring the effects of EEG signals on collision cases happening in the process of young drivers’ braking                                                                                                       & Research            & Stress               \\
\rowcolor[rgb]{0.753,0.753,0.753} \cite{Aspiotis2022}       & 2022                              & Assessing electroencephalography as a stress indicator: A VR high-altitude scenario monitored through EEG and ECG                                                                                                & Research            & Stress               \\
\cite{Cai2022}            & 2022                              & A multi-modal open dataset for mental-disorder analysis                                                                                                                                                          & Research            & Various              \\
\rowcolor[rgb]{0.753,0.753,0.753} \cite{Majid2022}          & 2022                              & A multimodal perceived stress classification framework using wearable physiological sensors                                                                                                                      & Research            & Stress               \\
\cite{Sharif2022}         & 2022                              & Evaluating the stressful commutes using physiological signals and machine learning                                                                                                                               & Research            & Stress               \\
\rowcolor[rgb]{0.753,0.753,0.753} \cite{AlShorman2022}      & 2022                              & Frontal lobe real-time EEG analysis using machine learning techniques for mental stress detection                                                                                                                & Research            & Stress               \\
\cite{Phutela2022}        & 2022                              & Stress classification using brain signals based on LSTM network                                                                                                                                                  & Research            & Stress               \\
\rowcolor[rgb]{0.753,0.753,0.753} \cite{Affanni2022}        & 2022                              & Development of an EEG headband for stress measurement on driving simulators                                                                                                                                      & Research            & Stress               \\
\cite{Hamatta2022}        & 2022                              & Genetic algorithm-based human mental stress detection and alerting in internet of things                                                                                                                         & Research            & Stress               \\
\rowcolor[rgb]{0.753,0.753,0.753} \cite{Sung2022}           & 2022                              & Real-time stress analysis affecting nurse during elective spinal surgery using a wearable device                                                                                                                 & Research            & Stress               \\
\cite{Wu2022}             & 2022                              & Emerging wearable biosensor technologies for stress monitoring and their real-world applications                                                                                                                 & Review              & Stress               \\
\rowcolor[rgb]{0.753,0.753,0.753} \cite{KumarGS2022}        & 2022                              & Wavelet based machine learning models for classification of human emotions using EEG signal                                                                                                                      & Research            & Emotion              \\
\cite{Rezaee2022}         & 2022                              & Fusion-based learning for stress recognition in smart home: An IoMT framework                                                                                                                                    & Research            & Stress               \\
\rowcolor[rgb]{0.753,0.753,0.753} \cite{Vanhollebeke2022}   & 2022                              & The neural correlates of psychosocial stress: A systematic review and meta-analysis of spectral analysis EEG studies                                                                                             & Review              & Stress               \\
\cite{Mohammed2023}       & 2023                              & Cheating detection in E-exams system using EEG signals                                                                                                                                                           & Research            & Stress               \\
\rowcolor[rgb]{0.753,0.753,0.753} \cite{PerezValero2023}    & 2023                              & Automated detection of Alzheimer's disease and other neurophysiological applications based on EEG                                                                                                                & Research            & Various              \\
\cite{Shermadurai2023} & 2023 & Deep learning framework for classification of mental stress from multimodal datasets & research & stress \\ 
\rowcolor[rgb]{0.753,0.753,0.753} \cite{Bhatnagar2023}      & 2023                              & A deep learning approach for assessing stress levels in patients using electroencephalogram signals                                                                                                              & Research            & Stress               \\
\cite{Suryawanshi2023a}   & 2023                              & Brain activity monitoring for stress analysis through EEG dataset using machine learning                                                                                                                         & Research            & Stress               \\
\rowcolor[rgb]{0.753,0.753,0.753} \cite{Caldiroli2023}      & 2023                              & Comparing online cognitive load on mobile versus PC-based devices                                                                                                                                                & Research            & Stress               \\
\cite{Sharif2023}         & 2023                              & An innovative random-forest-based model to assess the health impacts of regular commuting using non-invasive wearable sensors                                                                                    & Research            & Stress               \\
\rowcolor[rgb]{0.753,0.753,0.753} \cite{Lin2023}            & 2023                              & EEG emotion recognition using improved graph neural network with channel selection                                                                                                                               & Research            & Emotion              \\
\cite{Teo2023}            & 2023                              & Use of portable devices to measure brain and heart activity during relaxation and comparative conditions: Electroencephalogram, heart rate variability, and correlations with self-report psychological measures & Research            & Stress              \\
\label{tab:papersreviewed}
\end{longtable}
\end{landscape}

\FloatBarrier

\subsection{Assessment of the quality of the studies}
\noindent Two reviewers (Vos and Rahimi Azghadi) used the IJMEDI checklist \cite{Cabitza2021} to independently evaluate the included studies' quality. The IJMEDI checklist is a quality assessment tool for medical artificial intelligence studies proposed by the IJMEDI, which aims to distinguish high-quality machine learning studies from simple medical data-mining studies. Six dimensions are included as 30 questions in the checklist: problem and data understanding, data preparation, modeling, validation, and deployment. Each question can be answered as OK (adequately addressed), mR (sufficient but improvable), and MR (inadequately addressed). In high-priority items, OK, mR and MR were assigned scores of 2, 1, and 0, respectively, whereas, in low-priority items, the scores were halved. The maximum possible score was 50 points, with study quality divided into low (0–19.5), medium (20–34.5), and high (35–50).

\section{Results}

\subsection{Stress Assessment Using EEG}

\noindent The use of EEG in stress research relates to its purported sensitivity to localized brain activity in regions involved in the generation of the stress response, or activity associated with increased arousal or specific psycho-emotional states, with the majority of studies supporting the view that in a stressed state there is generally greater frontal right alpha activity in relation to the left alpha activity \cite{Giannakakis2019}. Cortisol response magnitude remains the gold standard indicator for stress assessment \cite{Boucher2019}, as cortisol levels quantify the endocrine response to stress \cite{Seo2010}. Seo \emph{et al.} \cite{Seo2010} reported a significant positive correlation between the cortisol level and relative high beta power at both the anterior temporal lobes (Figure \ref{fig:figure2}), and a tendency toward a similar relationship at one of the occipital sites. Barzegar \emph{et al.} \cite{Barzegar2023} similarly found salivary cortisol levels increased significantly along with the relative delta band frequency, while the beta bands and, in less amount, the theta and gamma decreased, especially in the frontal region. These effects on the cortisol stress hormone generally peak after about 20–30 min\cite{Westerink2020}. \\

\subsubsection{Sensor positions and recorded frequency bands}
\noindent Table \ref{tab:wavesreviewed} provides an overview of the studies incorporated in this review, explicitly detailing sensor positions and recorded frequency bands. Among these, five \cite{Pernice2020, Halim2020, Arsalan2021, Mohammed2023, Minguillon2016} focused on alpha waves, ten \cite{UmarSaeed2018, Saeed2020, Arsalan2021, Chae2021, Aspiotis2022, Majid2022, Mohammed2023, Sharif2023, Suryawanshi2023a, AlShorman2022} investigated beta waves, two \cite{Parent2020, Bhatnagar2023} explored delta waves, eight \cite{UmarSaeed2018, Saeed2020, Parent2020, Halim2020, Aspiotis2022, Minguillon2016, Suryawanshi2023a, AlShorman2022} examined gamma waves, and four \cite{Pernice2020, Halim2020, Majid2022, AlShorman2022} delved into theta frequency. Importantly, Newson \emph{et al.} \cite{Newson2019} in their review of EEG frequency bands in psychiatric disorders noted that it is not always obvious whether changes in particular frequency bands are specific to individual disorders, or whether overlap may exist, limiting clinical diagnosis potential. This finding is of further importance when machine learning classification models are utilized for diagnosis or prediction when such models are trained on datasets containing binary classes, for example, stressed/not-stressed. \\

\subsubsection{Data recording considerations}
\noindent Fifty of the 60 studies included in this review provided in-depth descriptive detail of the data that were used during experimentation. Of these, 35 (70\%) did not provide access to their study data, complicating the ability to easily reproduce study results. The remaining fifteen studies either published their datasets or used datasets previously made public. Additionally, 44\% of the total 60 studies reviewed did not explicitly note health screening before experimentation and data recording. A detailed health screening is of critical importance in order to build an accurate baseline dataset for training machine learning models, and as noted by Newson \emph{et al.} \cite{Newson2019}, underlying or undiagnosed mental health conditions can compromise model training if the data from such subjects are erroneously introduced into the baseline dataset.\\

\begin{landscape}
\fontsize{10}{11}\selectfont
\begin{longtable}{ C{1cm} C{1cm} p{8cm} p{1.5cm} p{1.5cm} p{1.5cm} p{1.5cm} p{1.5cm} }
\caption{Reported sensor placement with frequency band for stress studies included in this review.} \\
\hline\hline
\textbf{Paper} & \textbf{Year} & \textbf{Title} & \textbf{Frontal} & \textbf{Temporal} & \textbf{Central} & \textbf{Parietal} & \textbf{Occipital} \\
\hline
\hline\endhead  
\hline\endfoot  
\rowcolor[rgb]{0.753,0.753,0.753} \cite{Minguillon2016}     & 2016          & Stress assessment by prefrontal relative gamma                                                                                                            & Alpha, Beta, Theta & Alpha             & Alpha, Gamma     & Alpha             &                    \\
\cite{UmarSaeed2018}      & 2018          & Selection of neural oscillatory features for human stress classification with single channel EEG headset                                                  & Beta, Gamma        &                   &                  &                   &                    \\
\rowcolor[rgb]{0.753,0.753,0.753} \cite{Saeed2020}          & 2020          & EEG based classification of long-term stress using psychological labeling                                                                                 & Beta, Gamma        &                   &                  &                   &                    \\
\cite{Parent2020}         & 2020          & PASS: A multimodal database of physical activity and stress for mobile passive body/ brain-computer interface research                                    &                    & Gamma, Delta      &                  & Gamma, Delta      &                    \\
\rowcolor[rgb]{0.753,0.753,0.753} \cite{Pernice2020}        & 2020          & Multivariate correlation measures reveal structure and strength of brain-body physiological networks at rest and during mental stress                     & Theta, Alpha       &                   &                  &                   &                    \\
\cite{Halim2020}          & 2020          & On identification of driving-induced stress using electroencephalogram signals: A framework based on wearable safety-critical scheme and machine learning & Alpha              & Gamma             &                  & Theta, Alpha      & Gamma              \\
\rowcolor[rgb]{0.753,0.753,0.753} \cite{Arsalan2021}        & 2021          & Human stress classification during public speaking using physiological signals                                                                            & Alpha, Beta        &                   &                  &                   &                    \\
\cite{Chae2021}           & 2021          & Relationship between rework of engineering drawing tasks and stress level measured from physiological signals                                             & Beta               & Beta              &                  &                   &                    \\
\rowcolor[rgb]{0.753,0.753,0.753} \cite{Yeom2021}           & 2021          & Psychological and physiological effects of a green wall on occupants: A cross-over study in virtual reality                                               &                    &                   &                  &                   &                    \\
\cite{Aspiotis2022}       & 2022          & Assessing electroencephalography as a stress indicator: A VR high-altitude scenario monitored through EEG and ECG                                         & Beta               & Gamma, Beta       &                  & Gamma, Beta       & Gamma, Beta        \\
\rowcolor[rgb]{0.753,0.753,0.753} \cite{Majid2022}          & 2022          & A Multimodal perceived stress classification framework using wearable physiological sensors                                                               &                    & Theta, Beta       &                  & Theta, Beta       &                    \\
\cite{AlShorman2022}      & 2022          & Frontal lobe real-time EEG analysis using machine learning techniques for mental stress detection                                                         & Theta, Beta, Gamma &                   &                  &                   &            \\
\rowcolor[rgb]{0.753,0.753,0.753} \cite{Mohammed2023}       & 2023          & Cheating detection in E-exams system using EEG signals                                                                                                    & Alpha, Beta        &                   &                  & Alpha, Beta       &                    \\
\cite{Sharif2023}         & 2023          & An innovative random-forest-based model to assess the health impacts of regular commuting using non-invasive wearable sensors                             & Beta               &                   &                  &                   &                    \\

\rowcolor[rgb]{0.753,0.753,0.753} \cite{Bhatnagar2023}      & 2023          & A deep learning approach for assessing stress levels in patients using electroencephalogram signals                                                       & Delta              & Delta             &                  &                   &                    \\
\cite{Suryawanshi2023a}   & 2023          & Brain activity monitoring for stress analysis through EEG dataset using machine learning                                                                  & Beta, Gamma        &                   &                  &                   &                    \\

\label{tab:wavesreviewed}
\end{longtable}
\end{landscape}
        
\FloatBarrier

\subsection{Low-Cost EEG Devices in Stress Research}

\noindent Consumer-oriented EEG devices such as the Emotiv EPOC \cite{Emotiv} have been present in the market since at least 2010 (Table \ref{tab:devices}). Most of these devices would be considered low-cost in terms of total cost of ownership, operation and interpretation, given that they are mostly re-usable with supporting software and platforms to assist with interpretation of results. For this review, the scope for low-cost device inclusion was limited to those retailing for less than USD 1,000.00. Typical pricing for the devices included at the time of writing ranged from USD249.99 for the InteraXon Muse 2 (5 sensors) \cite{Muse}, up to USD849 for the Emotiv EPOC (14 sensors). This range was chosen to be comparable to smartwatches with embedded health monitoring sensors that can be used for stress measurement \cite{Vos2023}.

\subsubsection{Dominant low-cost EEG devices}
\noindent Of the studies included in this review using a low-cost EEG device, 23 used either the Emotiv \cite{Hou2015, Baghdadi2019, Saeed2020, Pernice2020, Caldiroli2023, Alakus2020, Halim2020, Chae2021, Yeom2021}, InteraXon \cite{Majid2022, Mohammed2023, Phutela2022, Parent2020, Arsalan2021, Sakalle2021} or NeuroSky \cite{Park2020, TuerxunWaili2020, Suryawanshi2023a, UmarSaeed2018, Sharif2023, Teo2023} range of devices \cite{NeuroSky} (Table \ref{tab:devices}). Two studies \cite{Sundaresan2021, Peterson2020} utilized the OpenBCI \cite{OpenBCI} series of devices consisting of an 8-channel OpenBCI Cyton board \cite{OpenBCI}, combined with a low-cost skull cap and electrodes. A lower cost Ganglion board with 4 channels is also available, retailing at USD499.00, compared to the Ganglion board which retails for USD999.00. OpenBCI introduced their EEG Headband Kit in 2022 which can be paired with either the Ganglion or Cyton board. The kit retails at USD279.99. \\

\subsubsection{Custom EEG devices}
\noindent Three studies included in this review used custom designed EEG devices, mostly built by hand. Lee \emph{et al.} \cite{Lee2020} designed a wearable in-and over-ear bio-metric sensing device that measured EEG and electrocardiogram (ECG) signals simultaneously. They further proposed a novel sensing electrode which is highly conductive, dry and flexible, and designed with portability and comfort in mind during long-term usage. Ahn \emph{et al.} \cite{Ahn2019} designed a wearable, two-channel EEG and single channel ECG device that was lightweight and exhibited excellent noise management performance. Affanni \emph{et al.} \cite{Affanni2022} designed a six-channel EEG wearable headband that transmits data over WiFi to a laptop, featuring a rechargeable battery with 10 hours of continuous transmission life.\\

\subsubsection{Additional EEG devices}
\noindent Fifteen studies \cite{Koelstra2012, Babayan2019, Aspiotis2022, Cai2022, PerezValero2023, Shermadurai2023, AlShorman2022, Minguillon2016, Hag2021, Hamatta2022, Mohanavelu2020, DiazPiedra2021, Tan2021, Zhang2021, KumarGS2022} provided detail on the EEG hardware used for recording. Of these, the BioSemi ActiveTwo \cite{BioSemi} device was the most prominent and was used in 7 studies. The remaining studies \cite{Lopez2015, Giannakakis2019, Newson2019, Katmah2021, Sharif2022, Bhatnagar2023, Giannakakis2015, Sung2022, Wu2022, Borghini2014, Hu2020, Zhang2020, Wan2021, Rezaee2022, Vanhollebeke2022} did not explicitly provide identifiable or named EEG hardware configuration. \\

\begin{table}[h!]
\centering
\caption{\label{tab:devices}Low-cost EEG devices included in this review.}
\resizebox{\textwidth}{!}{
\begin{tabular}{lccc} \\
\hline\hline
\textbf{Device}             & \textbf{Released} & \textbf{Sensors}      & \textbf{Data Download?}  \\
\hline
\rowcolor[rgb]{0.753,0.753,0.753} Emotiv EPOC                 & 2010                                      & AF3, AF4, F3, F4, F7, F8, FC5, FC6, O1, O2, P7, P8, T7, T8& Y                        \\
NeuroSky MindWave           & 2011                                      & FP1                                                           & Y                        \\
\rowcolor[rgb]{0.753,0.753,0.753}  NeuroSky Mindwave Mobile    & 2012                                      & FP1                                                           & Y                        \\
Emotiv EPOC+                & 2013                                      & AF3, AF4, F3, F4, F7, F8, FC5, FC6, O1, O2, P7, P8, T7, T8 & Y                        \\
\rowcolor[rgb]{0.753,0.753,0.753} InteraXon Muse                        & 2014                                      & AF7, AF8, Fpz, TP9, TP10                                      & Y                        \\
Emotiv INSIGHT              & 2015                                      & AF3, AF4, T7, T8, Pz                                          & Y                        \\
\rowcolor[rgb]{0.753,0.753,0.753} InteraXon Muse 2                      & 2018                                      & AF7, AF8, Fpz, TP9, TP10                                      & Y                        \\
NeuroSky MindWave Mobile II & 2018                                      & FP1                                                           & Y                        \\
\rowcolor[rgb]{0.753,0.753,0.753} OpenBCI EEG Headband Kit    & 2022                                      & Up to 8, User-configurable       & Y   \\
\hline 

\end{tabular}
}
\end{table}
\subsubsection{Sensor support and placement}
\noindent Given the wide range of frequency bands reportedly used for stress research, requiring multiple scalp sensors to cover multiple brain lobes (Table \ref{tab:wavesreviewed}), it is important to consider sensor support and placement for low-cost devices. As shown in Table \ref{tab:devices}, the Emotiv EPOC series of devices provide the broadest range of sensors (14), while the OpenBCI provides the most flexibility in sensor configuration due to its ability to pair with any sensor configuration (up to 16 channels). The NeuroSky Mindwave Mobile 2 device provides a single sensor resting on the forehead above the eye (FP1 position) with a separate reference electrode that is clipped to the left ear. Prior studies have reported the FP1 position to be involved in cognition, working memory and perception \cite{Bludau2013}. \\

\noindent In contrast, wrist and ring based wearable devices used for stress research typically offer a smaller, but more specialized range of sensors including for measuring heart or pulse rate, electrodermal activity, skin temperature and a three-axis accelerometer. This leads to more restrictive but simpler device placement and more standardization of sensor types across devices, resulting in less pre-processing during data analysis. Additionally, due to the nature of the recorded biomarkers, data produced by these sensors are simpler to standardize across devices requiring less pre-processing for reproducing prior study results.\\

\subsection{Datasets for EEG-based Stress Measurement}

\noindent Datasets that are publicly available (or available upon request) provide researchers with pre-labeled EEG sensor data for experimentation and reproducibility of prior results. Eight publicly available datasets were included in this review (Table \ref{tab:datasets}), of which three (PASS, BrainBodyStress, EDPMSC) \cite{Parent2020, Pernice2020, Arsalan2019} are specifically designed and labeled for stress. Four are labeled for emotion (DEAP, SEED, K-EmoCon, GAMEEMO) \cite{Koelstra2012, Zheng2015, Park2020, Alakus2020} and one dataset (DASPS) is labeled for anxiety \cite{Baghdadi2019}. Although labeled for emotion, the DEAP, SEED, K-Emocon and GAMEEMO datasets could potentially be utilized for stress research due to measuring states of arousal and valence which are known characteristics of emotional stress \cite{Christianson1992}. Among the datasets reviewed, the Emotiv EPOC EEG device was the most frequently used. The remaining datasets use a variety of other EEG devices.\\

\subsubsection{Dataset use and consideration}
\noindent Some of the eight datasets included in this review are used more frequently. For instance, six studies utilized the DEAP dataset \cite{Koelstra2012, Shermadurai2023, Hamatta2022, Tan2021, KumarGS2022, Lin2023} for experimentation, while only two used the SEED dataset \cite{Zheng2015, Lin2023}. Parent \emph{et al.} \cite{Parent2020}  designed and utilized the PASS stress dataset. Similarly, Alakus \emph{et al.} \cite{Alakus2020} designed and utilized the GAMEEMO stress dataset, while Pernice \emph{et al.} \cite{Pernice2020} developed the BrainBodyStress dataset. Arsalan \emph{et al.} \cite{Arsalan2019} collected and curated the EDPMSC dataset. The DEAP, DASPS, PASS, GAMEEMO, BrainBodyStress and EDPMSC datasets specifically noted employing a health screening protocol during dataset development.

\subsubsection{Incorporation of additional biomarkers}
\noindent 
Four datasets included additional biomarkers to enable extended experimentation (Table \ref{tab:datasets}), including ECG, Galvanic Skin Response (GSR) \cite{Koelstra2012, Parent2020, Pernice2020}, environmental Temperature (TEMP) \cite{Parent2020, Park2020}, HR \cite{Park2020}, Blood Volume Pulse (BVP) \cite{Park2020, Pernice2020}, Inter-Beat Interval (IBI) and EDA \cite{Park2020}. For non-EEG biomarkers, the device most commonly used for data recording was the Empatica E4 wrist-worn health monitor \cite{Empatica2022}, commonly used in stress research \cite{Siirtola2020, Gjoreski2017, Vos2023, Schuurmans2020}.

\subsubsection{Recording duration and stress labeling methods}
\noindent The recording duration (Table \ref{tab:datasets}) of the datasets included in this review varied from 3 minutes (DASPS dataset) to over 30 minutes (PASS and BrainBodyStress datasets). Stress is not a binary condition, further complicating clinical diagnosis when assessing EEG results. Researchers typically employ several different known methods for labeling or assessing biomarker data including the STROOP Color and Word Test \cite{Stroop1935, Hamid2019}, utilized in \cite{Hou2015, Bhatnagar2023}, the Perceived Stress Scale \cite{Cohen1983}, utilized in \cite{UmarSaeed2018, Saeed2020, Majid2022}, and the State-Trait Anxiety Inventory \cite{Spielberger1983}, utilized in \cite{Halim2020, Phutela2022}. Alternatively, researchers may label biomarker data according to relaxation periods and periods where stress is induced, as utilized in \cite{Parent2020, Pernice2020, Mohammed2023, Teo2023}.\\

\noindent The DEAP dataset \cite{Koelstra2012} recorded a 2-minute non-stressed baseline followed by a 1-minute display of a music video to assess arousal, valence, liking and dominance. This was repeated for 20 trials. The SEED dataset \cite{Zheng2015} provided a 5-second hint prior to displaying a movie for 4 minutes, followed by a 45-second period to allow for self-assessment via a questionnaire. This was followed by a 15-second rest period, repeated for a total of 15 trials. The DASPS datasets \cite{Baghdadi2019} performed a total of 6 trials consisting of a 5-minute baseline, 30-second stimuli, and a 4-minute post-stimuli period during which a self-assessment questionnaire was completed. \\

\noindent The PASS dataset \cite{Parent2020} simulated a stressful event through the use of two video games, one being stressful while the other was considered non-stressful. Additionally, this dataset implemented a 2-minute baseline followed by 5 to 15 minutes of game-play, with a 5-minute break between sessions. Two datasets provided short, but well-balanced periods of baseline and stress conditions. The BrainBodyStress dataset \cite{Pernice2020} used a 12-minute relaxation baseline, followed by a 12-minute sustained attention task and a 7-minute mental arithmetic task, while the EDPMSC dataset \cite{Arsalan2019} used a 3-minute relaxation baseline followed by 3 minutes of public speaking, and a 3-minute post-activity phase.

\subsubsection{Data formats and accessibility}
\noindent The datasets included in this review are provided in a wide range of file formats, including Comma-Separated Values (CSV), MATLAB \cite{FormatMATLAB}, and the European Data Format (EDF) \cite{FormatEDF}. Researchers looking to utilize these datasets in their studies are therefore required to use statistical software capable of supporting these formats or develop external conversion scripts. A number of these datasets provided both raw and pre-processed filtered data. Providing raw EEG data, as is the case with the DEAP, K-EmoCon, PASS and GAMEEMO datasets (Table \ref{tab:datasets}), allows researchers to experiment with various pre-processing and filtering techniques to match their experimental requirements. 

\subsubsection{Data processing and artifact removal}
\noindent Low-cost, consumer-oriented health tracking and monitoring devices frequently provide a sophisticated software platform with built-in analytics to deliver insights directly to users, often as a paid subscription model. While this is of great benefit to consumers and other non-experts, for researchers this likely implies less control over data processing and artifact removal. The NeuroSky range of devices transmit processed data via a wireless Bluetooth connection to a computer, and offer a wide range of plug-in modules to extend the base software functionality of the device. NeuroSky devices apply a Fast Fourier Transform (FFT) to the raw EEG prior to transmission. For the Emotiv EPOC series devices, baseline removal is applied prior to finite impulse response (FIR) filtering, channel rejection and independent component analysis (ICA) for artifact removal. The InteraXon Muse devices can be configured to provide raw, unprocessed data, while the OpenBCI devices provide raw EEG signals by default. \\

\noindent Pernice \emph{et al.} \cite{Pernice2020} performed artifact removal using a high-pass filter (half power frequency of 1 Hz) and a low-pass filter (half power frequency of 20 Hz), followed by power spectral density (PSD) calculation for each EEG signal using a 2-second sliding window with 50\% overlap. Hou \emph{et al.} \cite{Hou2015}, Suryawanshi \emph{et al.} \cite{Suryawanshi2023a} and AlShorman \emph{et al.} \cite{AlShorman2022} applied FFT to extract the mean power spectrum for EEG frequency bands (delta, theta, alpha, beta and gamma), while Majid \emph{et al.} \cite{Majid2022} applied FFT using a window size of 256 with an overlap of 90\%. Before applying the FFT, AlShorman \emph{et al.} \cite{AlShorman2022} filtered the raw signal by using a band pass filter with a cut-off frequency of 0.1–30 Hz, while Suryawanshi \emph{et al.} \cite{Suryawanshi2023a} applied a band-pass filter at 1Hz, subsequently followed by a low-pass filter at 50Hz for artifact removal, a standard practice in EEG signal processing \cite{Subramaniyam2023}.\\

\noindent While artifact removal and filtering techniques are common across all studies reviewed, and are a normal part of EEG analysis, slight differences in algorithmic application and sequence of implementation could potentially hinder the reproducibility of prior reported results \cite{Katmah2021, Safayari2021, Newson2019, Wu2022}; more so when the EEG signals are recorded using different devices. Reproducibility would therefore require the exact step-by-step pre-processing techniques, statistical software, libraries and algorithms on the raw EEG sensor data. \\

\noindent Numerous factors can affect EEG measurement \cite{Wu2022} including sensor type, placement, signal processing methods and artifact treatment and removal. In this review, a large and varied number of methods were found to be utilized showing a lack of standardization and no definitive approach to signal processing employed across the studies included in this review. Katmah \emph{et al.} \cite{Katmah2021}, Safayari \emph{et al} \cite{Safayari2021} and Newson \emph{et al.} \cite{Newson2019} noted critical differences reported with respect to model accuracy and input features between stress-related studies utilizing EEG and suggested the variation could be due to lack of standardization during data pre-processing, feature extraction and experimental setup and duration. Newson \emph{et al.} \cite{Newson2019} further noted differences in band ranges that define each frequency range due to either hardware configuration and limitations or pre-processing implementation during data analysis, ultimately leading to differences in results interpretation.\\

\begin{landscape}
\fontsize{10}{11}\selectfont
\begin{longtable}{ p{1cm} p{1cm} p{2.7cm} p{3.2cm} p{2cm} C{1.3cm} C{1cm} C{1cm}  C{1cm}  p{1.5cm} p{1.5cm}}
\caption{Datasets included in this review.} \\
\hline\hline
\textbf{Paper} & \textbf{Year} & \textbf{Dataset} & \multicolumn{1}{l}{\textbf{Device(s)}}                 & \textbf{Biomarkers}               & \textbf{Subjects} & \textbf{Gender} & \textbf{Age (Mean)} & \textbf{Health} & \textbf{Condition} & \textbf{Duration} \\
\hline
\hline\endhead  
\hline\endfoot  

\rowcolor[rgb]{0.753,0.753,0.753} \cite{Koelstra2012}       & 2012                              & DEAP             & Biosemi ActiveTwo                  & EEG, GSR, ECG                     & 32    & 16 F, 16 M
 & 26.9                               & Y                         & Emotion (SAM)      &  15 min   \\
\cite{Zheng2015}          & 2015                              & SEED             & NeuroScan                          & EEG                               & 15   & 8 F, 7 M
 &                               &                & Emotion           & 5 min\\
\rowcolor[rgb]{0.753,0.753,0.753} \cite{Baghdadi2019}       & 2019                              & DASPS            & Emotiv EPOC                        & EEG                               & 23      & 13 F, 10 M
 & 30                           & Y                         & Anxiety (HAM-A, SAM)        &  3 min \\
\cite{Arsalan2019}            & 2019                              & EDPMSC           & InteraXon Muse                       & EEG        & 40 & 20 F, 20 M
 & 24.85                                   & Y                         & Stress            & 6 min \\
\rowcolor[rgb]{0.753,0.753,0.753} \cite{Parent2020}         & 2020                              & PASS             & InteraXon Muse, BioHarness 3, E4             & EEG, ECG, EDA, TEMP               & 48       &     &                         & Y                         & Stress (NASA-TLX, BORG)            & 30 min\\
\cite{Park2020}           & 2020                              & K-EmoCon         & NeuroSky MindWave, E4, Polar H7    & EEG, HR, TEMP, IBI, BVP, EDA   & 32  & 12 F, 20 M
 & 23.8                                  &                 & Emotion       &  10 min   \\
\rowcolor[rgb]{0.753,0.753,0.753} \cite{Alakus2020}         & 2020                              & GAMEEMO          & Emotiv EPOC+                       & EEG                               & 28     & & 23.5
                             & Y                         & Emotion (SAM)          & 20 min \\
\cite{Pernice2020}        & 2020                              & BrainBodyStress  & Emotiv EPOC+, Empatica E4         & EEG, ECG, BVP                     & 18   & 5 F, 13 M
 & 25                                & Y                         & Stress        &  30 min  

\label{tab:datasets}
\end{longtable}
\end{landscape}
\FloatBarrier

\subsection{Machine Learning Techniques for EEG-based Stress Measurement}

\noindent Of the studies included in this review, thirty applied machine learning techniques to predict or measure acute stress response. Of these, thirteen utilized a low-cost EEG device to record brain activity, as shown in Table \ref{tab:mlreviewed}. The study by AlShorman \emph{et al.} \cite{AlShorman2022} where a 128-channel EEG device was used, and Bhatnagar \emph{et al.} \cite{Bhatnagar2023} study did not report their devices. Both of these studies reported the highest accuracy rates from the thirty studies reviewed and were therefore used as a baseline for comparison to low-cost EEG results reported.\\

\subsubsection{Data modalities}
\noindent Ten of the fifteen studies utilized only EEG data as input to their machine learning algorithms. Five studies \cite{Parent2020, Pernice2020, Arsalan2021, Majid2022, Sharif2023} implemented a multi-modal approach incorporating additional biomarker data from ECG or other health monitoring devices (EDA, GSR, TEMP, BP, PPG). Majid \emph{et al.} \cite{Majid2022} used a signal fusion approach to combine raw EEG from a Muse device with GSR and Photoplethysmography (PPG) signals from additional devices into a single feature set. 
Additionally, three studies \cite{Majid2022, Shermadurai2023, Lee2020} employed fusion-based architectures to combine the features from EEG with other health monitoring devices. \\

\subsubsection{Machine learning algorithms}
\noindent The most widely applied machine learning algorithm across the studies reviewed was the Support Vector Machine (SVM) (Table \ref{tab:mlreviewed}). SVM is a linear model commonly used for classification and regression problems in machine learning studies and can solve both linear and non-linear problems by creating a line or a hyperplane to separate input data into separate classes. When applied to stress research, these classes would typically be one of stressed or non-stressed. A number of more advanced techniques were noted in the studies reviewed, however, these were not applied to data recorded from low-cost EEG devices. Bhatnagar \emph{et al.} \cite{Bhatnagar2023} utilized EEGNet, a compact convolutional neural network for EEG-based brain–computer interfaces first proposed by Lawhern \emph{et al.} \cite{Lawhern2018}.  Also noted were the use of Spiking Neural Networks (SNN) \cite{Tan2021}, graph models \cite{Lin2023} and the use of genetic algorithms \cite{Hamatta2022}.\\

\begin{landscape}
\fontsize{10}{10}\selectfont
\begin{longtable}{ C{1cm} p{1cm} p{5cm} p{3cm} p{3cm} p{2.2cm} C{1.5cm} p{2cm}}
\caption{Machine Learning studies included in this review.} \\
\hline\hline
\textbf{Paper} & \textbf{Year} & \textbf{Device(s)} & \textbf{Biomarkers} & \textbf{Algorithm(s)} & \textbf{Accuracy} & \textbf{Subjects} & \textbf{Validation} \\
\hline
\hline\endhead  
\hline\endfoot  

\rowcolor[rgb]{0.753,0.753,0.753} \cite{Hou2015}           & 2015                              & Emotiv EPOC                         & EEG                 & SVM                   & 85.71                                 & 9                                     & 5-Fold            \\
\cite{UmarSaeed2018}      & 2018                              & NeuroSky MindWave                   & EEG                 & SVM                   & 78.57                                 & 28                                    & 10-Fold           \\
\rowcolor[rgb]{0.753,0.753,0.753} \cite{Halim2020}          & 2020                              & Emotiv EPOC+                       & EEG                 & SVM, NN, RF           & 97.95 (ENS)                                & 86                                    & 70/30                \\
\cite{Parent2020}         & 2020                              & InteraXon Muse, BioHarness 3, Empatica E4     & EEG, ECG, EDA, TEMP & SVM                   & 74                                    & 48                                    & 5-Fold, LOSO      \\
\rowcolor[rgb]{0.753,0.753,0.753} \cite{Pernice2020}        & 2020                              & Emotiv EPOC+, Empatica E4          & EEG, ECG, BVP       & LR                    &                                       & 18                                    & Fisher F-test        \\
\cite{Saeed2020}          & 2020                              & Emotiv INSIGHT                      & EEG                 & SVM, NB, KNN, LR, MLP & 85.2 (SVM)                                 & 33                                    & 10-Fold           \\
\rowcolor[rgb]{0.753,0.753,0.753} \cite{Arsalan2021}        & 2021                              & InteraXon Muse                                & EEG, GSR, PPG       & KNN, DT, RF, MLP, SVM & 96.25 (SVM)                                 & 40                                    & LOSO                 \\
\cite{Sundaresan2021}     & 2021                              & OpenBCI                             & EEG                 & SVM, CNN, LTSM        & 93.27 (LSTM)                                & 13                                    & 70/30                \\
\rowcolor[rgb]{0.753,0.753,0.753} \cite{AlShorman2022}	& 2022	& Geodesic	& EEG &	SVM, NB	 & 98 (SVM)	& 14 &	3-Fold CV \\

\cite{Majid2022}          & 2022                              & InteraXon Muse Headband, Shimmer GSR          & EEG,GSR, PPG        & SVM, NB, MLP          & 95 (MLP)                                    & 40                                    & LOSO                 \\
\rowcolor[rgb]{0.753,0.753,0.753} \cite{Phutela2022}        & 2022                              & InteraXon Muse Headband                       & EEG                 & MLP, LSTM             & 93.17 (LTSM)                                & 40                                    & 10-Fold           \\
\cite{Bhatnagar2023}      & 2023                              &                                     & EEG                 & EEGNet                & 99.45                                 & 45                                    & 10-Fold           \\
\rowcolor[rgb]{0.753,0.753,0.753} \cite{Mohammed2023}       & 2023                              & InteraXon Muse 2                              & EEG                 & CNN                   & 97.37                                 & 15                                    & 70/30                \\
\cite{Sharif2023}         & 2023                              & NeuroSky Mindwave Mobile, MySignals & EEG, BP             & RF, KNN, SVM, NB      & 91 (RF)                                   & 45                                    & 10-Fold           \\
\rowcolor[rgb]{0.753,0.753,0.753}  \cite{Suryawanshi2023a}   & 2023                              & NeuroSky Mindwave Mobile            & EEG                 & SVM, KNN, NN      & 90 (SVM)                                    & & 70/30 
\label{tab:mlreviewed}
\end{longtable}

\end{landscape}

\FloatBarrier

\noindent As detailed in Table \ref{tab:mlreviewed}, nine studies tested several different algorithms, with the best-performing algorithm shown in the \emph{Accuracy} column. SVM outperformed other algorithms in all but four studies. Of those, two \cite{Sundaresan2021, Phutela2022} utilized a dual-layer Long Short-Term Memory (LSTM) network. The study by Sharif \emph{et al.} \cite{Sharif2023} was the only one where a tree-based algorithm provided the highest accuracy of any combination. While Halim \emph{et al.} \cite{Halim2020} achieved their highest accuracy by combining the outputs of SVM, Random Forest (RF) and a Neural Network (NN) through an ensembling (ENS) approach, SVM outperformed both RF and NN before combining the predictions. \\

\noindent Eight of the fifteen studies \cite{Hou2015, UmarSaeed2018, Parent2020, Saeed2020, AlShorman2022, Phutela2022, Bhatnagar2023, Sharif2023} validated their results using a K-Fold cross-validation approach, while four \cite{Halim2020, Sundaresan2021, Mohammed2023, Suryawanshi2023a} utilized a more traditional train-test split, where 70\% of the data was used for training, and the predicted results validated against the remaining 30\%. \\

\noindent Of the fifteen studies listed in Table \ref{tab:mlreviewed}, seven reported the relative importance of the frequency bands used as algorithm input features. Three studies \cite{Halim2020, Pernice2020, Bhatnagar2023} reported the delta band to be the most important input feature, while alpha was reported by two \cite{Pernice2020, Arsalan2021}, beta by two \cite{Arsalan2021, Sharif2023}, gamma by two \cite{Halim2020, Parent2020} and theta by a single study \cite{Majid2022}. Frontal and temporal lobes were the most reported locations for sensor placement \cite{Parent2020, Pernice2020, Bhatnagar2023}. \\ 

\subsection{Study Quality}

\noindent The \emph{Supplementary File} details the results of the IJMEDI quality assessment. Table \ref{tab:ijmedi} summarizes the scores of each dimension and the total score of each study. The average score of the included studies was 26 (range: 11–42.5). Most of the studies were of medium quality, while two \cite{Parent2020, Saeed2020} were of high quality.\\

\begin{figure}[h!]
\centering
\includegraphics[width=\textwidth]{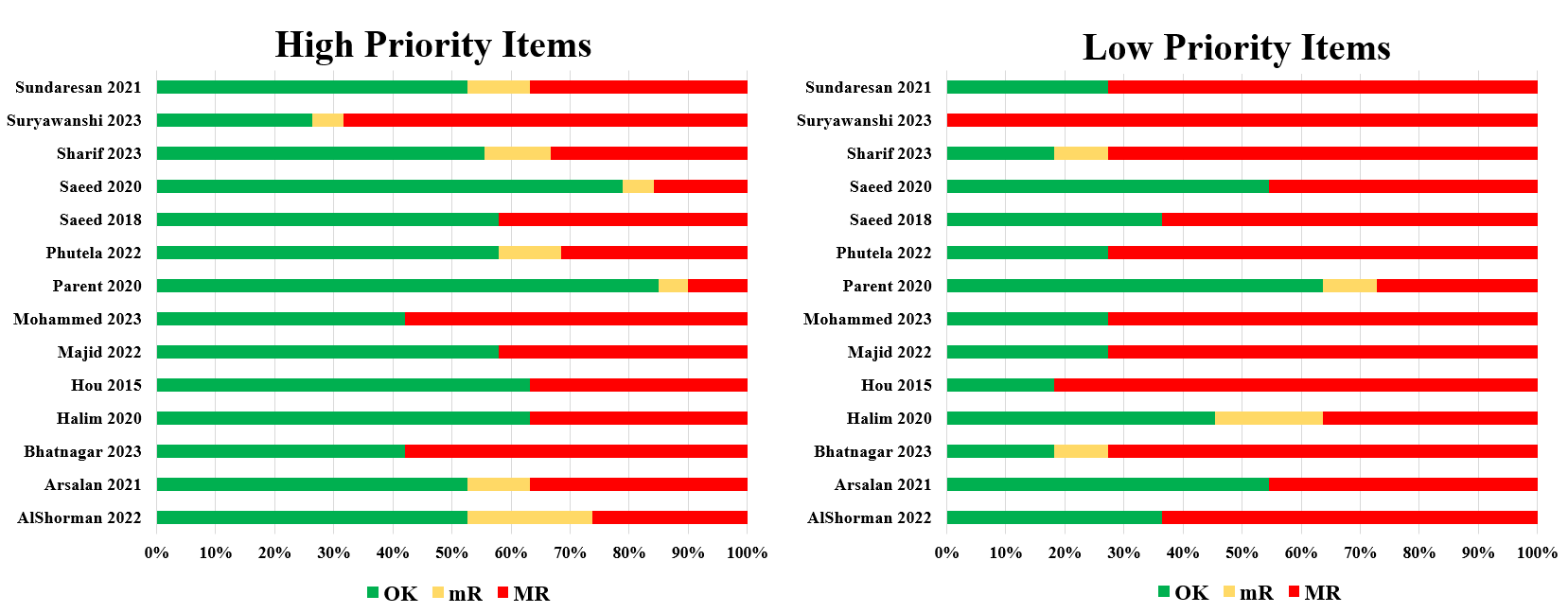}
\caption{\label{fig:figure6} Study quality assessment based on the IJMEDI checklist. Proportion of the different answers in the high- and low-priority items. OK = adequately addressed; mR = sufficient but improvable; MR = inadequately addressed.}%
\end{figure}

\noindent The majority of the studies lacked quality in the data preparation, validation and deployment dimensions. Figure \ref{fig:figure6} shows the proportion of the different answers in the high and low priority items. Importantly, high-priority items were generally well addressed, with clear problem understanding and a very strong focus on machine learning modeling techniques and implementation.\\

\begin{sidewaystable}
\centering
\caption{Quality assessment scores of the 14 ML-based studies according to the IJMEDI checklist.}
\renewcommand{\arraystretch}{1.5}
\resizebox{\textwidth}{!}{
\begin{tabular}{lcccccccc}
\\
\hline\hline
                   & \textbf{Problem Understanding (10)} & \textbf{Data Understanding (6)} & \textbf{Data Preparation (8)} & \textbf{Modeling (6)} & \textbf{Validation (12)} & \textbf{Deployment (8)} & \textbf{Total (50)} \\
\hline

\rowcolor[rgb]{0.753,0.753,0.753} \cite{AlShorman2022}    & 8                                                       & 5                                                   & 4                                                 & 6                                         & 5                                            & 1                                           & 29                                       \\
\cite{Arsalan2021}     & 7                                                       & 5                                                   & 2                                                 & 6                                         & 6                                            & 3                                           & 29                                       \\
\rowcolor[rgb]{0.753,0.753,0.753} \cite{Bhatnagar2023}    & 5                                                       & 0                                                   & 0                                                 & 6                                         & 7                                            & 0.5                                         & 18.5                                     \\
\cite{Halim2020}        & 8                                                       & 4                                                   & 2                                                 & 6                                         & 7.5                                          & 2.5                                         & 30                                       \\
\rowcolor[rgb]{0.753,0.753,0.753} \cite{Hou2015}          & 7                                                       & 4                                                   & 2                                                 & 6                                         & 7                                            & 1                                           & 27                                       \\
\cite{Majid2022}        & 8                                                       & 2                                                   & 2                                                 & 6                                         & 7                                            & 0                                           & 25                                       \\
\rowcolor[rgb]{0.753,0.753,0.753} \cite{Mohammed2023}     & 5                                                       & 0                                                   & 0                                                 & 6                                         & 7                                            & 1                                           & 19                                       \\
\cite{Parent2020} & 9 & 4 & 8 & 6 & 9 & 6.5 & 42.5 \\

\rowcolor[rgb]{0.753,0.753,0.753} \cite{Phutela2022}      & 8                                                       & 4                                                   & 1                                                 & 6                                         & 7                                            & 0                                           & 26                                       \\
\cite{UmarSaeed2018}   & 8                                                       & 2                                                   & 2                                                 & 6                                         & 6                                            & 1                                           & 25                                      \\
\rowcolor[rgb]{0.753,0.753,0.753} \cite{Saeed2020}       & 10                                                      & 6                                                   & 4                                                 & 6                                         & 8                                            & 2                                           & 36                                       \\
\cite{Sharif2023}      & 7.5                                                     & 2                                                   & 4                                                 & 6                                         & 6                                            & 0                                           & 25.5                                     \\
\rowcolor[rgb]{0.753,0.753,0.753} \cite{Sundaresan2021} & 8 & 4 & 3 & 6 & 6 & 0 & 27 \\

\cite{Suryawanshi2023a} & 2                                                       & 0                                                   & 1                                                 & 6                                         & 2                                            & 0                                           & 11                                       \\

\hline\hline
\end{tabular}
}
\label{tab:ijmedi}
\end{sidewaystable}
\FloatBarrier

\section{Discussion}

\noindent In order to build a robust machine learning model capable of accurately measuring stress via low-cost EEG sensors, we consider three important requirements. These include (i) EEG devices need to be technically capable of reading signals that correlate with acute stress response; (ii) For classifying stress through machine learning, high-quality training data needs to be available to properly train a classification algorithm; (iii) The resulting trained machine learning model should generalize by accurately predicting for new, unseen data. The discussion of this review is therefore focused on those three key requirements.

\subsection{Requirement 1: Low-cost EEG Device Capabilities}
\begin{figure}[h!]
\centering
\fbox{\includegraphics[width=\textwidth]{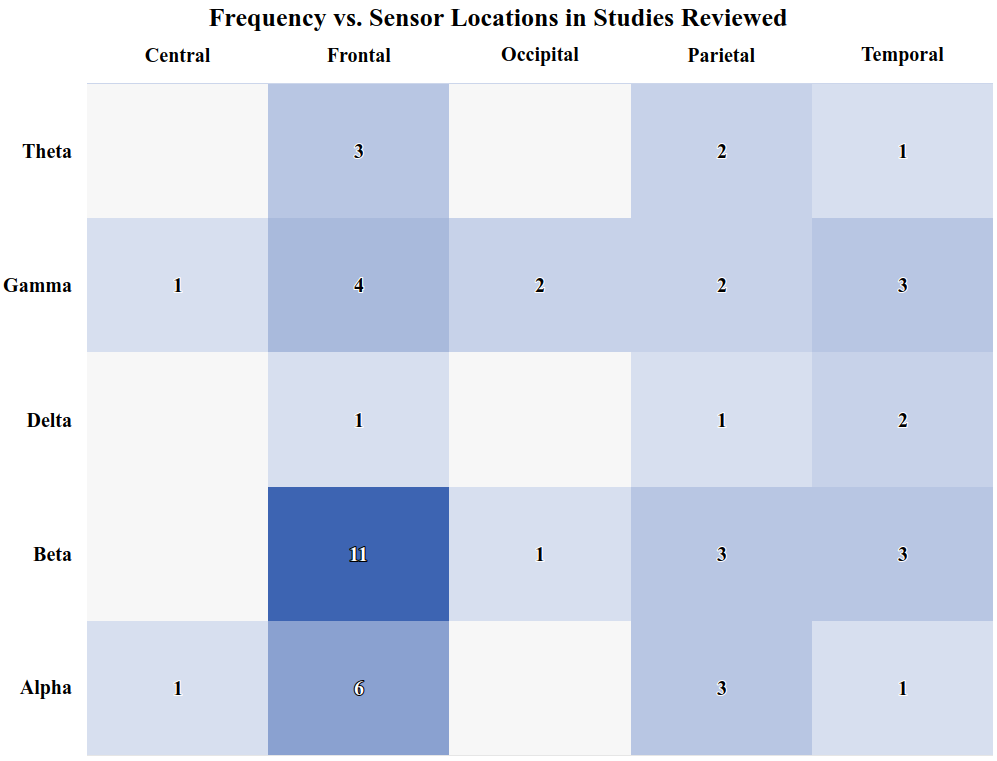}}
\caption{\label{fig:figure7}Frequency band utilization and sensor locations of the low-cost EEG devices reviewed in this study.}%
\end{figure}

\noindent Low-cost, wireless EEG devices are generally limited in terms of sensor placement and count (Table \ref{tab:devices}). Hence, experiments utilizing these devices would be naturally biased towards reporting positive outcomes concerning these scalp locations. The majority of the studies included in this review focused on the frontal, parietal and temporal regions (Figure \ref{fig:figure7}), likely due to limited device sensor support (Table \ref{tab:devices}) for central and occipital sites. Within these studies, the beta, gamma and alpha frequency bands provided the highest correlation with stress. \\

\FloatBarrier

\subsubsection{Sensor support}
\noindent Based on the findings of these aforementioned studies, these devices do appear to offer sufficient sensor support for measuring stress. Seo \emph{et al.} \cite{Seo2010} reports close relationships among EEG, ECG, and salivary cortisol indicative of acute stress, with chronic stress particularly associated with high levels of relative beta power, albeit at anterior temporal sites. Both the Emotiv and InteraXon series of devices does have sensor support across frontal, parietal and temporal areas, with the Emotiv range additionally supporting occipital sites.

\subsubsection{Ease of use}
\noindent Of further importance to researchers when opting to use low-cost, wireless EEG devices are ease of use \cite{UmarSaeed2018, Halim2020, Majid2022, Mohammed2023} and acquisition and operational costs \cite{Phutela2022, Mohammed2023}. A number of studies included in this review noted the mobility and affordability of EEG headsets as a key requirement for their specific study requirements \cite{UmarSaeed2018, Halim2020, Majid2022, Mohammed2023, Sharif2023}. Halim \emph{et al.} \cite{Halim2020} performed a study to evaluate the feasibility of detecting stress while subjects were driving motor vehicles, while Arsalan \emph{et al.} \cite{Arsalan2021} performed a study on measuring stress during public speaking. A mobile EEG device was therefore a technical necessity for both studies. However, very few studies have investigated stress detection and measurement in the presence of varying levels of physical activity \cite{Parent2020}. \\

\subsubsection{Technical challenges}
\noindent None of the studies included in this review reported any technical challenges, device failure or significant signal quality issues that required additional care or action during experimentation. While there are numerous benefits in using these lower cost EEG devices, researchers need to be acutely aware of any limitations when using such devices. For example, Pernice \emph{et al.} \cite{Pernice2020} noted in their study that EEG devices with dry electrode, scalp level connectivity, such as those reviewed in this study, do not permit a perfectly reliable interpretation of interacting brain areas, as they can be corrupted by volume conduction effects or confounding factors. However, Hinrichs \emph{et al.} \cite{Hinrichs2020} compared dry and wet EEG systems and found slightly higher artefacts for dry EEG systems while being more robust to 50Hz line noise, and concluded that dry EEG systems comply with the needs of clinical applications. \\

\subsection{Requirement 2: Machine Learning Input Data Quality}

\noindent Robust machine learning models require high-quality training data, and a number of datasets are publicly available containing EEG data for experimentation and/or model validation (Table \ref{tab:datasets}). Three of these datasets are specifically labeled for stress, of which the PASS dataset contains the largest sample size and includes additional biomarkers (ECG, EDA, TEMP) that have previously been shown to be useful features for building machine learning models capable of detecting and measuring stress \cite{Seo2010, Minguillon2016, Vos2023}.

\subsubsection{Data variability}
\noindent A key challenge in utilizing these datasets for validation of trained models would be adjustment for differences in device brand and sensor availability. Variation in signal quality or sensor type could hinder a direct data comparison when using both datasets as training and or validation pairs \cite{Wu2022}, while differentiation in experimental recording protocol could introduce noise into the data \cite{Wu2022}. In their review, Katmah \emph{et al.} \cite{Katmah2021} highlighted critical differences between research findings and argued that variations in the data analysis methods contributed to several contradictory results. Newson \emph{et al.} \cite{Newson2019} advise to refrain from making any clinical interpretations based on current findings due to the significant lack of standardization in frequency band definition.\\

\noindent Researchers considering the use of these datasets, whether for training purposes or for the validation of novel models, will consequently need to perform substantial pre-processing while accounting for potential device limitations in order to ensure robust and accurate analyses.

\subsubsection{Sample size}
\noindent Of the datasets included in this review, most contained data for a small sample size of test subjects, recorded using a number of different EEG devices. Campos-Ugaz \emph{et al.} \cite{CamposUgaz2023} noted that almost all EEG studies have small sample sizes, which considerably reduces the generalizability of their findings. Sundaresan \emph{et al.} \cite{Sundaresan2021} noted high inter- and intra-individual variability of neuro-physiological signals to be a central challenge for EEG-based stress studies, and further suggested using datasets containing a sufficiently large number of test subjects.\\

\noindent However, simply increasing sample size may not address the issue of generalization alone. Chaumon \emph{et al.} \cite{Chaumon2021} used Magnetoencephalography (MEG) data from a large sample of subjects to examine the behavior of statistical power measures and concluded that there is no unequivocal answer to the question of an appropriate sample size in MEG or EEG experiments. Boudewyn \emph{et al} \cite{Boudewyn2017} specifically aimed to determine the number of trials required to achieve an Event-Related Potential (ERP) effect in EEG studies, and concluded that several factors have an impact on statistical power in ERP studies, including number of trials, sample size, and effect magnitude, as well as interactions among these factors.

\subsubsection{Data labeling}
\noindent Finally, a critical aspect when performing supervised classification using machine learning approaches is accurate data labeling. This refers to the process of marking segments of data to correspond with the expected class, for example stressed or non-stressed. The datasets included in this review all included a baseline period labeled as non-stressed, with an experimental period marked as stressed. The duration of both baseline and experimental periods differed notably across the datasets, with the baseline period often being short, resulting in a class imbalance that researchers will need to adjust for when using these datasets for classification tasks.

\subsubsection{Study quality}
\noindent Having scored the machine learning studies included in this review using the IJMEDI checklist, we found two \cite{Parent2020, Saeed2020} of high quality, with nine studies being of medium quality \cite{AlShorman2022, Arsalan2021, Halim2020, Hou2015, Majid2022, Phutela2022, Sharif2023, Sundaresan2021, UmarSaeed2018}, and two studies being of low quality \cite{Mohammed2023, Suryawanshi2023a}. Most studies scored well in problem, validation and modeling dimensions. Data preparation and deployment scores were notably low, with little focus on the deployment of models in real-life scenarios, including factors pertaining to sustainability, model bias and ethics. For data preparation scoring, the IJMEDI checklist focuses on four key areas including outlier detection and treatment, class balancing, missing-value identification and management and data pre-processing, and lower scores indicate that these items are not described or described in sufficient detail. The highest overall quality score was attained by Parent \emph{et al.} \cite{Parent2020} (42.5), with high scores across all dimensions. The study by Pernice \emph{et al.} \cite{Pernice2020} was excluded from the assessment due to it being primarily focused on multivariate correlation analysis to investigate dynamic interactions between organ systems in the human body during stress.\\

\subsection{Requirement 3: Machine Learning Model Generalization}

\begin{figure}[h!]
\centering
\fbox{\includegraphics[width=\textwidth]{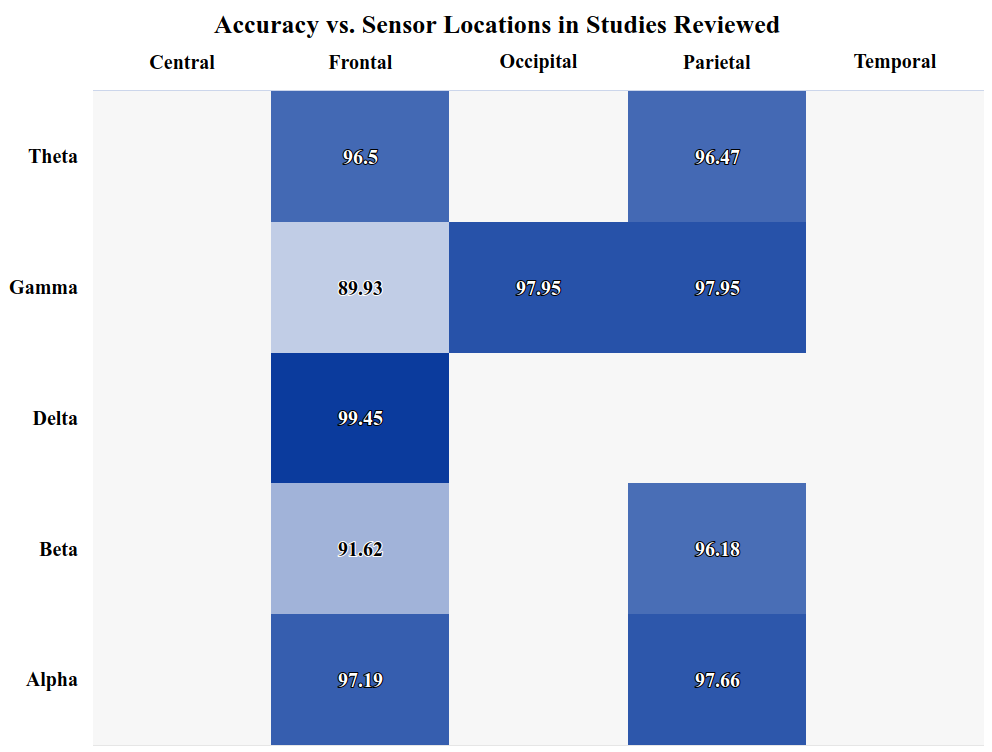}}
\caption{\label{fig:figure8}Reported accuracy by frequency band utilization and sensor locations.}%
\end{figure}

\noindent When training a machine learning model, researchers need to select the appropriate learning method based on the desired outcome. This outcome could be binary (stressed or non-stressed) or contiguous (level of stress being experienced). All studies reviewed developed a binary model. 
In addition, supervised and unsupervised learning \cite{Soni2020} are two different types of machine learning techniques which differ in the way the models are trained, and the type of training data required. All fifteen studies included in this review utilized supervised training and reported prediction for a binary class (stressed/non-stressed). Figure \ref{fig:figure8} provides a summary of the mean predictive accuracy reported across these studies, grouped by EEG sensor location and frequency band used as primary algorithm input features.

\subsubsection{Prediction accuracy and its significance}
\noindent Ten of the fifteen studies (66.6\%) reported predictive accuracy of more than 90\% \cite{Halim2020, Arsalan2021, Sundaresan2021, AlShorman2022, Majid2022, Phutela2022, Bhatnagar2023, Mohammed2023, Sharif2023, Suryawanshi2023a}, with the frontal and parietal regions providing the most consistent results across the gamma, alpha and beta frequency bands. However, Newson \emph{et al.} \cite{Newson2019} highlighted in their examination of EEG frequency bands within psychiatric disorders that distinguishing whether alterations in specific frequency bands are unique to individual disorders or if there might be overlaps, poses challenges that could restrict the potential for clinical diagnosis. 

\subsubsection{Health screening}
\noindent Of the datasets included in this review, three \cite{Arsalan2019, Zheng2015, Park2020} did not provide extensive detail on health screening of subjects included in their datasets, while a further three noted that subjects were healthy \cite{Koelstra2012, Alakus2020, Pernice2020}. Two datasets \cite{Baghdadi2019, Parent2020} specifically noted that test subjects were health screened specifically for neurological and psychological disorders. The lack of consistent health screening and detailed information on the screening process could potentially affect study reproducibility, and bring into question the high accuracy results reported should undiagnosed mental health conditions be present in the data. \\

\FloatBarrier

\subsubsection{Validation methods}
\noindent A number of methods were employed for results validation in the studies reviewed (Table \ref{tab:mlreviewed}), including self-scoring questionnaires used as correlation measures against the predicted stress level reported by the machine learning algorithm. Saeed \emph{et al.} \cite{Saeed2020, UmarSaeed2018} used the Perceived Stress Score, while Halim \emph{et al.} \cite{Halim2020} reported using the Self-Assessment Manikins, devised by Bradley \emph{et al.} \cite{Bradley1994}. Phutela \emph{et al.} \cite{Phutela2022} used the State Trait Anxiety Inventory. Statistical methods were also reported, with Parent \emph{et al.} \cite{Parent2020} validating their findings using a two-way ANOVA test, while Pernice \emph{et al.} \cite{Pernice2020} used the Fisher F-Test. Luck \emph{et al.} \cite{Luck2016} however noted that statistically significant but unpredicted and unreplicable effects in EEG experiments can easily be found due to the high random variation in EEG data, making it hard for researchers to determine if a noted effect is real, and that study replication and reproducibility should be key to experiment validation. None of the fifteen studies reviewed validated their findings on a new, unseen dataset.\\

\subsubsection{Cross validation using other wearables}
\noindent Three studies that incorporated signals from external devices to augment their input features \cite{Arsalan2021, Majid2022, Sharif2023} noted a significant correlation between heart rate and the predicted level of stress. Seo \emph{et al.} \cite{Seo2010} specifically reported a correlation between HRV and high beta activity at anterior temporal sites, opening the possibility of combining EEG with other wearable devices such as the Empatica range to cross-validate stress measurement between these devices. While a number of studies included in this review \cite{Teo2023,Parent2020, Pernice2020, Minguillon2016} did measure BVP, HR and HRV, none specifically cross-validated stress reported via EEG with those reported via heart rate. This could potentially provide a way forward for researchers by using additional wearable devices, such as the Empatica \cite{Empatica2022} range, to record heart rate and other biomarkers that have been shown to correlate with increased levels of stress \cite{Samson2020}, and using those as a validation mechanism for stress-related EEG studies. \\

\subsection{Summary}
\noindent The significant observations from this review are:
\begin{itemize}
  \item Low-cost EEG devices are increasingly being used for mental health related studies, including stress. However, sensor availability and placement differ widely across devices, potentially limiting the reproducibility of results reported from machine learning studies utilizing these devices.
  \item A small number of datasets are publicly available for researchers to use in stress related studies, including for results validation. However, most studies included in this review did not validate their findings and results on an independent, unseen dataset.
  \item None of the studies included in this review sufficiently considered the impact of underlying, undiagnosed mental health conditions on model predictive accuracy. This is especially important when considering the small size of study groups within the studies included in this review.
\end{itemize}

\subsection{Challenges and future research directions}
\noindent To achieve robust and reliable machine learning models suitable for real-world monitoring of stress using low-cost EEG devices, future research should address three formidable challenges:
\begin{itemize}
  \item The pre-processing of EEG data needs to be standardized to enable cross-device studies with the same machine learning models. Future research can develop a standard framework for dealing with data produced by low-cost EEG devices.
  \item While several EEG datasets recorded using low-cost devices have been made public, future work can provide larger datasets containing more significant and high-quality baseline (non-stressed) periods along with stressed periods to reduce data imbalance and assist researchers with model training and validation.
  \item A limited number of the studies reviewed utilized additional biomarkers recorded with non-EEG wearable devices. The use of other wearable devices that record biomarkers including heart rate and galvanic skin response could be further explored in two different ways. One is as a mechanism for validating the findings from stress-related studies, where EEG signals are the primary data-providing mechanism. Another is in a multi-modal approach, where both EEG and wearable biomarkers are used for stress prediction.
\end{itemize}

\section{Conclusion}
\noindent The main objective in algorithmic stress detection and measurement is to develop a robust, highly accurate model that can provide reproducible results. Low-cost EEG devices can potentially facilitate the development of machine learning models that produce accurate stress measurements and are reproducible across studies, irrespective of the study group or study setting. Nonetheless, several important factors should be taken into account. The review presented here synthesized the literature and discussed these essential factors in previous studies. In particular, we reviewed and analyzed the devices available, and their use in stress-related studies using machine learning techniques, while including their advantages, limitations and experimental quality. We also summarized our point of view on the challenges and opportunities in this emerging domain. We believe this review will advance knowledge in the general area of machine learning for stress detection using low-cost EEG devices, helping the research efforts move one step closer to realizing effective stress detection and management technology.

 \bibliographystyle{elsarticle-num} 
 \bibliography{cas-refs}





\end{document}